\documentclass[showpacs]{revtex4}
\def\ii{{\rm i}}
\def\nn{\nonumber}
\def\pa{\partial}
\usepackage{epsfig,amsmath,amssymb}
\begin{document}
%%%%%%%%%%%%%%%%%%%%%%%%%%%%%%%%%%%%%%%%%%%%%%
\title{A new approach to the study of quasi-normal modes of rotating stars}

\author{Valeria Ferrari$^{1,2}$, Leonardo Gualtieri$^{1,2}$, Stefania
Marassi$^{1,2}$} 

\affiliation{ $^1$ Dipartimento di Fisica ``G. Marconi", Universit\`a
``La Sapienza", I-00185 Roma, Italy\\ $^2$ INFN, Sezione di Roma,
I-00185 Roma, Italy }

\begin{abstract}
We propose a new method to study the quasi-normal modes of rotating
relativistic stars. Oscillations are treated as perturbations in the
frequency domain of the stationary, axisymmetric background describing
a rotating star. The perturbed quantities are expanded in circular
harmonics, and the resulting 2D-equations they satisfy are integrated
using spectral methods in the $(r,\theta)$-plane.  The asymptotic
conditions at infinity, needed to find the mode frequencies, are
implemented by generalizing the standing wave boundary condition
commonly used in the non rotating case.  As a test, the method is
applied to find the quasi-normal mode frequencies of a slowly rotating
star.
\end{abstract}
\maketitle

%%%%%%%%%%%%%%%%%%%%%%%%%%%%%%%%%%%%%%%%%%%%%%
\section{Introduction}\label{intro}
%%%%%%%%%%%%%%%%%%%%%%%%%%%%%%%%%%%%%%%%%%%%%%
Non radial oscillations of compact stars can be excited in several
astrophysical events.  For instance, after a neutron star (NS) is
formed in a gravitational collapse, or in processes that may occur
during its subsequent evolution; these include starquakes, glitches,
interactions with a stellar companion, or phase transitions to quark
matter or to a kaon and/or pion condensate, that may arise in the
inner core of a NS if the density exceeds some critical value.  All
these phenomena induce perturbations which set the star in oscillation
and, according to general relativity, gravitational waves (GWs) are
one of the channels through which energy is dissipated.

In addition, due to rotation some modes may grow unstable through the
Chandrasekhar-Friedman-Schutz mechanism (CFS instability) \cite{CFS};
these instabilities may have important effects on the subsequent
evolution of the star, and they may be associated to a further
emission of GWs, the amount of which would depend on when and whether
the growing modes are saturated by non-linear couplings or dissipative
processes.  It is therefore important to know at which frequencies a
star pulsates emitting gravitational waves, and to study under which
conditions the corresponding modes become unstable.

If one assumes that the star does not rotate, the mode frequencies can
easily be computed by solving the equations of stellar perturbations,
which have been formulated in the Sixties \cite{thorne1} and further
developed in later years \cite{chandrafer,chandrafer1}. These
equations have been integrated for a large variety of equations of
state proposed to describe matter in a NS \cite{AS}.  These studies
show that the identification of the frequency corresponding to the
excitation of a stellar mode (for instance the fundamental mode which
is likely to be the most energetic) in a detected gravitational
signal, would allow to infer interesting information on the
composition of the inner core of a NS and on the equation of state of
matter at supranuclear densities.

However, all stars rotate, and our present knowledge of the
quasi-normal mode (QNM) spectrum of rotating stars is far to be
complete.  The perturbative approach which works so fine in the non
rotating case, when generalized to include rotation shows a high
degree of complexity, even if the star is only slowly rotating
\cite{Kojima1992,Kojima1993,KojimaPTP,RSK,VB,FGPS,VBH,PGMF,SPK,LFA}.
A major difficulty arises because, when using the standard spherical
harmonics decomposition of the perturbed tensors, modes with different
harmonic indexes couple, giving rise to an infinite set of dynamical,
coupled equations.  For this reason, in all studies based on this
approach simplifying assumptions are introduced: either the couplings
between oscillations with different values of the harmonic index $l$
are neglected, or Cowling's approximation is used (i.e. spacetime
perturbations are neglected)
\cite{Kojima1993,KojimaPTP,RSK,VB,FGPS,VBH,PGMF,SPK}.  To our
knowledge, the only place where the oscillations of a slowly rotating
star are studied without making use of any of these restrictive
assumptions is in \cite{LFA} where, however, only $r$-modes have been
considered.

An alternative approach consists in solving the equations describing a
rotating and oscillating star in full general relativity, in time
domain.  However, current studies based on this approach also make use
of strong simplifying assumptions, or restrict to particular cases.
For instance, the Cowling approximation has been used in several
papers \cite{Cowling}; in \cite{Fal} only quasi-radial modes ($l=0$)
have been considered; in \cite{SF,MSB} only the neutral mode
(zero-frequency mode in the rotating frame) has been studied; in
\cite{DSF} only axisymmetric ($m=0$) modes have been analysed, using
the conformal flatness condition.  In \cite{FDGS}, the frequencies of
axisymmetric modes ($m=0$) with $l=0$ to $l=3$ have been computed for
rapidly rotating relativistic polytropes, using the Cowling
approximation; a comparison of the results for $l=0$ obtained in the
Cowling approximation in \cite{FDGS}, with those found in full GR in
\cite{Fal}, shows that Cowling's approximation introduces large errors
in the determination of the fundamental mode frequency.

In this article we develop a general method to find the quasi-normal
mode frequencies of a rotating star. We perturb Einstein's equations
about a stationary, axisymmetric background describing a rotating
star. The perturbed quantities are expanded in circular harmonics
$e^{\ii m\phi}$. As we are looking for quasi-normal modes, we assume a
time dependence $e^{-\ii\omega t}$, with $\omega$ complex.  Due to the
background symmetry, perturbations with different values of $\omega$
and $m$ are decoupled; thus, for assigned values of $(\omega,m)$, the
perturbed equations to solve are a 2D-system of linear, differential
equations in $r$ and $\theta$.  In this paper we do not derive
explicitly the perturbed equations in the general case of rapidly
rotating stars, since they will be studied in a subsequent paper. Our
goal here is to describe the general method.

Two are the main ingredients on which our method is based: i) the
perturbed equations are integrated using spectral methods; ii) the
boundary conditions at the center of the star and at radial infinity
are implemented by suitably generalizing the standing wave approach
which has been used to find the QNM frequencies of non-rotating stars
\cite{SW,SW2,chandrafer,chandrafer1}. These points will be described
in Section \ref{method}.  To test our approach, in Section
\ref{slowly} we find the frequency of the fundamental mode of a slowly
rotating, constant density star, as a function of the rotation
rate. Concluding remarks are in Section \ref{conclusions}.
%%%%%%%%%%%%%%%%%%%%%%%%%%%%%%%%%%%%%%%%%%%%%%%%%%%%%%%%%%%%%%%
\section{The general  method}\label{method}
%%%%%%%%%%%%%%%%%%%%%%%%%%%%%%%%%%%%%%%%%%%%%%%%%%%%%%%%%%%%%%%
The metric which we consider as a background is stationary and axially
symmetric. It can be cast in the general form \cite{Bardeen}
\begin{equation}
(ds^2)^{(0)}=g^{(0)}_{\mu\nu}dx^\mu dx^\nu=e^{-\nu}dt^2+
e^\psi(d\phi-\omega dt)^2+e^{\mu_2}dx_2^2+e^{\mu_3}dx_3^2\label{rotmet1}
\end{equation}
where $\nu$, $\psi$, $\mu_2$, $\mu_3$ are functions of the coordinates
$x_2$, $x_3$. In  (\ref{rotmet1}) there is still some gauge
freedom, which allows to write the metric in a simpler form, like
\begin{equation}
(ds^2)^{(0)}=e^{-\nu(r,\theta)}dt^2+
e^{\psi(r,\theta)}(d\phi-w(r,\theta) dt)^2+e^{\mu(r,\theta)}
(dr^2+r^2d\theta^2)\label{rotmet2}
\end{equation}
as in \cite{SF}, or like
\begin{equation}
(ds^2)^{(0)}=e^{-\nu(r,\theta)}dt^2+
e^{\lambda(r,\theta)}dr^2+e^{\mu(r,\theta)}r^2\left(d\theta^2
+\sin^2\theta(d\phi-w(r,\theta) dt)^2\right)\label{rotmet3}
\end{equation}
as in \cite{HS}, \cite{Hartle}.  In the following we shall not specify
explicitly the gauge, but we will require that some properties are
satisfied; in particular we require: i) that the spacetime is
described by the coordinates $(t,r,\theta,\phi)$, ii) that
$\frac{\partial}{\partial t}$, $\frac{\partial}{\partial\phi}$ are
Killing vectors associated with stationarity and axisymmetry
respectively, and iii) that $\theta,\phi$ are polar coordinates on
spheres, i.e. that the surfaces $t=const$, $r=const$ are diffeomorphic
(but not isomorphic) to $2$-spheres.  As a consequence of these
assumptions (that are fulfilled by (\ref{rotmet2}) and
(\ref{rotmet3})) any tensor field defined on one of these surfaces can
formally be expanded in tensor spherical harmonics, even if the
spacetime is not spherically symmetric; this property will be useful
in Section \ref{standingrot}.

The metric and fluid velocity perturbations can be considered as
tensor fields in this background; they are expanded in circular
harmonics $e^{\ii m\phi}$, and Fourier-transformed in time.  Since
perturbations belonging to different $m$ and different frequency
$\omega$ do not couple, in what follows $m$ and $\omega$ will be
considered as fixed, and perturbed quantities will be decomposed as
follows
\begin{eqnarray}
h_{\mu\nu}(t,r,\theta,\phi)&=&h^{m\,\omega}_{\mu\nu}(r,\theta)
e^{\ii m\phi}e^{-\ii\omega t}\nn\\
\delta u_{\mu}(t,r,\theta,\phi)&=&\delta u^{m\,\omega}_{\mu}(r,\theta)
e^{\ii m\phi}e^{-\ii\omega t}\nn\\
\delta\rho(t,r,\theta,\phi)&=&\delta\rho^{m\,\omega}(r,\theta)
e^{\ii m\phi}e^{-\ii\omega t}\nn\\
\delta p(t,r,\theta,\phi)&=&\delta p^{m\,\omega}(r,\theta)
e^{\ii m\phi}e^{-\ii\omega t}\,.\label{hurp}
\end{eqnarray}
The frequency $\omega$ is, in general, complex.

By fixing the gauge, imposing $u^\mu u_\mu=-1$ and assigning an
equation of state which relates $\delta p$ and $\delta\rho$, the
sixteen quantities $h^{m\,\omega}_{\mu\nu}$, $\delta
u^{m\,\omega}_\mu$, $\delta\rho^{m\,\omega}$, $\delta p^{m\,\omega}$
reduce to ten $\{H_i^{m\,\omega}(r,\theta)\}_{i=1,\dots,10}$.  A
possible gauge choice, which we call {\it generalized Regge-Wheeler
gauge}, is described in Appendix \ref{genRW}; however, other gauges
can be considered (see for instance \cite{SF}). For simplicity of
notation, to hereafter we shall assume that the quantities
$H_i^{m\,\omega}$ are scalars with respect to rotation (as they are,
indeed, in the generalized Regge-Wheeler gauge, see Appendix
\ref{genRW}); however, every step of the approach we will describe can
be applied also to tensorial quantities by a suitable generalization.

Einstein's equations, linearized about our stationary axisymmetric
background, reduce to a system of partial differential equations (PDE)
for the functions $H_i^{m\,\omega}$  in
$r$ and $\theta$. To find the QNM frequencies, for assigned
values of $m$ and $\omega$, we solve these equations by
imposing that all metric functions are regular near the center of the
star, that the Lagrangian perturbation of the pressure vanishes on the
stellar surface, and that the solution at infinity behaves as a pure
outgoing wave. The conditions at the center and at the surface of the
star can be fulfilled for every value of $\omega$, but the outgoing
wave condition at infinity is only consistent with a discrete set of
(complex) frequencies $\{\omega_i\}$; such frequencies are the QNM.

We will now describe how to implement the boundary condition at
infinity, and the numerical approach we use to solve the 
perturbed equations.
%%%%%%%%%%%%%%%%%%%%%%%%%%%%%%%%%%%%%%%%%%%%%%
\subsection{The boundary condition at radial infinity}\label{standingwave}
%%%%%%%%%%%%%%%%%%%%%%%%%%%%%%%%%%%%%%%%%%%%%%
In this section we shall generalize to
the rotating case the {\it standing wave} approach \cite{SW,SW2} used to find
the QNM frequencies of non rotating stars. To this
purpose, it is useful to remind how this method works.

%%%%%%%%%%%%%%%%%%%%%%%%%%%%%%%%%%%%%%%%%%%%%%%%%%%%%%%%%%%%%%%%%%%%%%%%%%%%%
\subsubsection{The standing wave approach for spherical stars}
%%%%%%%%%%%%%%%%%%%%%%%%%%%%%%%%%%%%%%%%%%%%%%%%%%%%%%%%%%%%%%%%%%%%%%%%%%%%%
First of all it is worth stressing that by this approach
\cite{SW,SW2} one can only  determine the QNM frequencies of
slowly damped modes, like the $f$-, $p$-, and $r$-modes;  it cannot
be applied to highly damped modes, like stellar $w$-modes or
black hole's QNM.

It is known that the equations describing the perturbations of a non
rotating, spherical star can be separated by expanding the perturbed
tensors in tensorial spherical harmonics; outside the star these
equations reduce to those describing Schwarzschild perturbations, and
they can be reduced to the Regge-Wheeler \cite{RW} and the Zerilli
\cite{Zer} equations, for two suitably defined functions which we both
indicate as $Z^{lm}(r,\omega)$.  The two wave equations have the
following form
\begin{equation}
\frac{d^{2}Z^{lm}(r,\omega)}{dr^{2}_{*}}+\left[\omega^{2}-V(r) \right]
Z^{lm}(r,\omega)=0,\qquad\quad l\ge 2~,
\label{rwzer}
\end{equation}
where $r_*$ is the usual tortoise coordinate and $V(r)$ is a short
range potential.  The QNM frequencies are the values of the complex
frequency $\omega$ for which the solutions of Equation (\ref{rwzer}), found
by imposing appropriate boundary conditions at the surface of the
star, behave as pure outgoing waves at radial infinity, i.e.
\begin{equation}
Z^{lm}(r,\omega)=
A^{lm}_{out}(\omega)e^{\ii\omega r_*}\qquad\quad
\hbox{as} \qquad\quad r_* \longrightarrow \infty\,.
\label{defainout}
\end{equation}
The standing wave approach consists in the following.  Let us assume
that $Z^{lm}(r,\omega)$ is an analytic function of the complex
variable $\omega=\sigma-\ii/\tau$, and be
$\omega_0=\sigma_0-\ii/\tau_0$ the frequency of a QNM, with
$|1/\tau_0|\ll\sigma_0$.  In general, at radial infinity the solution
of Equation (\ref{rwzer}) is a superposition of ingoing and outgoing waves,
i.e.
\begin{equation}
Z^{lm}(r,\omega)=A^{lm}_{in}(\omega)e^{-\ii\omega r_*}+
A^{lm}_{out}(\omega)e^{\ii\omega r_*}\,.\label{realZ}
\end{equation}
If $\omega= \omega_0$, by definition $A^{lm}_{in}(\omega)=0$;  since $Z^{lm}$ is analytic and
since $|1/\tau_0|\ll\sigma_0$, we can expand $A^{lm}_{in}(\sigma)$ near the
real $\sigma_0$ as follows
\begin{equation}
A^{lm}_{in}(\omega_0)=A^{lm}_{in}(\sigma_0)-\frac{\ii}{\tau_0}
A^{lm\,\,\prime}_{in}(\sigma_0)~,
\end{equation}
where ``$~'~$'' indicates differentiation with respect to $\sigma$;
then, by imposing $A^{lm}_{in}(\omega_0)=0$ we find
\begin{equation}
A^{lm\,\,\prime}_{in}(\sigma_0)=-\ii\tau_0 A^{lm}_{in}(\sigma_0)\,.
\end{equation}
Using this relation the function $A^{lm}_{in}(\sigma)$, near
$\sigma=\sigma_0$ (with $\sigma$ and $\sigma_0$ real), can be written
as
\begin{eqnarray}
A^{lm}_{in}(\sigma)&=&A^{lm}_{in}(\sigma_0)+(\sigma-\sigma_0)
A^{lm\,\,\prime}_{in}
(\sigma_0)= - \ii\tau_0A^{lm}_{in}(\sigma_0)\left[(\sigma-\sigma_0)
+\frac{\ii}{\tau_0}\right]\,,
\label{parcompl}
\end{eqnarray}
from which it follows
\begin{equation}
\left|A^{lm}_{in}(\sigma)\right|^2
=B^2\left[(\sigma-\sigma_0)^2+\frac{1}{\tau_0^2}\right]~,
\label{stanwav}
\end{equation}
where $B$ is a constant which does not depend on $\sigma$.  Thus, to
find the frequencies of the QNM it is sufficient to integrate the
wave equation (\ref{rwzer}) for {\it real} values of the frequency
$\sigma$, and find the values $\sigma_{i}$ for which the amplitude
of the standing wave (\ref{stanwav}) has a minimum: these are the QNM
frequencies. The corresponding damping times $\tau_i$ can be found 
through a quadratic fit of $\left|A^{lm}_{in}(\sigma)\right|^2$.
%%%%%%%%%%%%%%%%%%%%%%%%%%%%%%%%%%%%%%%%%%%%%%%%%%%%%%%%%%%%%%%%%%%%%%%%%%%%%
\subsubsection{The standing wave approach for rotating stars}
\label{standingrot}
%%%%%%%%%%%%%%%%%%%%%%%%%%%%%%%%%%%%%%%%%%%%%%%%%%%%%%%%%%%%%%%%%%%%%%%%%%%%%
Let us now consider a rotating star.  As discussed above, with a
suitable choice of the gauge the relevant perturbed quantities reduce
to a set of quantities which behave as scalars with respect to
rotation: $\{H_i^{m\,\omega}(r,\theta)\}_{i=1,\dots,10}$ ($\omega$
complex). They must satisfy a set of PDE, obtained by linearizing
Einstein's equation, which can be integrated once the values of these
quantities are assigned at the center of the star, i.e. on a sphere of
radius $r_0\ll R$ (hereafter $R$ is the stellar radius)
\begin{equation}
H_i^{m\,\omega}(r_0,\theta)=H_{0i}^{m\,\omega}(\theta)
\,.\label{condcent}
\end{equation}
The $H_{0i}^{m\,\omega}(\theta)$ are subject to constraints, which
arise from the analytical expansion in powers of $r$ of the perturbed
equations, from the assumption of regularity of the spacetime as
$r\rightarrow 0$, and from the requirement that the Lagrangian
pressure perturbation must vanish on the surface of the star.  These
constraints reduce the number of independent quantities from the ten
$H_{0i}^{m\,\omega}(\theta)$ to a smaller number, say $N$,
i.e. $\{\hat H_{0j}^m(\theta)\}_{j=1,\dots,N}$.  Being these
quantities defined on a sphere $r=r_0$, they can formally be
decomposed in spherical harmonics
\begin{equation}
\hat H_{0j}^m(\theta)=\sum_{l=|m|}^L\hat H_j^{lm}Y^{lm}(\theta,0)\,,
\label{Hexpans}
\end{equation}
where  the expansion is truncated  at $l=L$.  Therefore the
independent solutions of the perturbed equations correspond to the
following set of $N\cdot\left[L-|m|+1\right]$ constants
\begin{equation}
\left\{\hat H_j^{lm}\right\}
\qquad\hbox{with}\qquad
j=1,\dots , N\qquad\hbox{and}\qquad
l=|m|,\dots,L\,.
\label{Hconst}
\end{equation}
Given these constants,  the perturbed equations for the functions
$H_i^{m\,\omega}(r,\theta)$  can be integrated for $r\geq r_0$.

In the wave zone, far away from the star,  the far field limit expansion of
the metric describing a rotating star shows that the metric
reduces to the Schwarzschild solution 
(see for instance \cite{MTW}, Chap. 19). This occurs because
terms due to rotation decrease faster than the ``Schwarzschild-like'' components.
Therefore, as when dealing with Schwarzschild perturbations, 
in this asymptotic region we can define the gauge invariant Zerilli and
Regge-Wheeler functions, $Z_{Zer}^{lm}(r,\omega)$ and $Z_{RW}^{lm}(r,\omega)$,
in terms of the  perturbed metric tensor, expanded in tensorial spherical harmonics with $l \ge 2$.  
This tensor is found by integrating the equations describing 
the perturbed spacetime outside the rotating star.
The well known  asymptotic behaviour of $Z_{Zer}^{lm}(r,\omega)$ and
$Z_{RW}^{lm}(r,\omega)$ is
\begin{eqnarray}
Z_{Zer}^{lm}(r,\omega)&=&A^{lm}_{Zer\,in}(\omega)e^{-\ii\omega r_*}+
A^{lm}_{Zer\,out}(\omega)e^{\ii\omega r_*}\nn\\
Z_{RW}^{lm}(r,\omega)&=&A^{lm}_{RW\,in}(\omega)e^{-\ii\omega r_*}+
A^{lm}_{RW\,out}(\omega)e^{\ii\omega r_*}\,.
\end{eqnarray}
A (complex) frequency $\omega_0$ belongs to a quasi-normal mode if,
for an assigned value of $m$, the following condition is satisfied for
{\it any} $l$:
\begin{equation}
A^{lm}_{Zer\,in}(\omega_0)=
A^{lm}_{RW\,in}(\omega_0)=0
~~~\forall l\label{eqmoderot0}
\end{equation}
i.e. if the set of $2\cdot\left[L-|m|+1\right]$ constants 
\begin{equation}
\left\{A^{lm}_{Zer\,in}(\omega),A^{lm}_{RW\,in}(\omega)\right\}
\qquad\hbox{with}\qquad
l=|m|,\dots,L
\label{Aconst}
\end{equation}
vanishes.  

It should be stressed that this is a big difference with respect to
the non rotating case: in that case each mode belongs to a single,
assigned value of $l$, and there is degeneracy in $m$.

For each assigned value of $m$, we define the
vectors
\begin{equation}
{\bf\hat H}^m\equiv
\left(\begin{array}{c}
\hat H_1^{|m|\,m}\\
\hat H_1^{|m|+1\,m}\\
\vdots\\
\\
\hat H_2^{|m|\,m}\\
\hat H_2^{|m|+1\,m}\\
\vdots\\
\\
\end{array}\right)~~~~~\hbox{and}~~~~~
{\bf A}^m\equiv
\left(\begin{array}{c}
A_{Zer\,in}^{|m|\,m}(\omega)\\
A_{Zer\,in}^{|m|+1\,m}(\omega)\\
\vdots\\
\\
A_{RW\,in}^{|m|\,m}(\omega)\\
A_{RW\,in}^{|m|+1\,m}(\omega)\\
\vdots\\
\\
\end{array}\right)
\end{equation}
the dimensionality of which is $N\cdot\left[L-|m|+1\right]$ and
$2\cdot\left[L-|m|+1\right]$,
respectively.  Since the perturbed equations are linear, these vectors
are related by the matrix equation
\begin{equation}
{\bf A}^m(\omega)={\bf M}^m(\omega){\bf\hat H}^m\,;
\end{equation}
the constants $\hat H_j^{lm}$ do not depend on $\omega$.
The coefficients of the complex matrix ${\bf M}^m(\omega)$ 
have to be evaluated by integrating the perturbed equations.

Equation (\ref{eqmoderot0}), which identifies the QNM
eigenfrequencies, can be written as
\begin{equation}
{\bf M}^m(\omega_0){\bf \hat H}^m=0~~\forall{\bf \hat H}^m\,.
\label{eigeneqn}
\end{equation}
A discrete set of QNM's exists if  the matrix ${\bf M}$ is square, i.e. if $N=2$. 
Thus, equation (\ref{eigeneqn}) is equivalent to impose 
\begin{equation}
\det\left({\bf M}^m(\omega_0)\right)=0\,.\label{eqmoderot}
\end{equation}
As we will show in the Appendices, by counting the number of
independent equations in the cases of spherical stars and of slowly
rotating stars we find indeed $N=2$. We expect the same to hold also
for rapidly rotating stars.

Let us now restrict the frequency to the real axis. Furthermore, we
normalize the constants ${\bf\hat H}^m$ so that the solutions
$H_i^{m\,\sigma}(r,\theta)$ (and consequently
$Z_{Zer}^{lm}(r,\sigma)$, $Z^{lm}_{RW}(r,\sigma)$) are real; this is
always possible, because the perturbed Einstein equations in the
frequency domain have real coefficients, as long as $\sigma$ is real
(if we assume that the fluid is non dissipative, so that the equations
in time domain are time-symmetric).  Thus, in the wave zone we have
\begin{eqnarray}
Z_{Zer}^{lm}(r,\sigma)&=&A^{lm}_{Zer\,in}(\sigma)e^{-\ii\sigma r_*}+
A^{lm}_{Zer\,out}(\sigma)e^{\ii\sigma r_*}~\in~I\!\!R\nn\\
Z_{RW}^{lm}(r,\sigma)&=&A^{lm}_{RW\,in}(\sigma)e^{-\ii\sigma r_*}+
A^{lm}_{RW\,out}(\sigma)e^{\ii\sigma r_*}~\in~I\!\!R\,.
\end{eqnarray}
The ingoing  wave amplitudes, $A^{lm}_{Zer\,in}$ and $A^{lm}_{RW\,in}$, can be found, as shown in 
\cite{SW2}, by evaluating $Z_{Zer}$ and $Z_{RW}$ at different values of $r_*$, fitting
the two functions as a superposition of $\sin \sigma r_*$ and $\cos \sigma r_*$.
It should be noted  that although ${\bf\hat H}$ are real, the quantities
$A^{lm}_{Zer\,in}$, $A^{lm}_{RW\,in}$ are complex (they satisfy the
conditions $A^{lm}_{Zer\,out}=(A^{lm}_{Zer\,in})^*$,
$A^{lm}_{RW\,out}=(A^{lm}_{RW\,in})^*$), thus ${\bf M}$ is complex.  
For $\sigma$ real, the vectors ${\bf\hat H}^m$, ${\bf A}^m$ are related by
\begin{equation}
{\bf A}^m(\sigma)={\bf M}^m(\sigma){\bf\hat H}^m\,.
\end{equation}
By expanding equation (\ref{eqmoderot}) about
$\omega_0=\sigma_0-\ii/\tau_0$ (with $|1/\tau_0|\ll\sigma_0$) as we
did for spherical stars, we find that if
$\sigma\sim\sigma_0$
\begin{equation}
\det{\bf M}^m(\sigma)=\det{\bf M}(\sigma_0)[1-\ii\tau_0
(\sigma-\sigma_0)]~~~~~\Rightarrow~~~~~
\left|\det{\bf M}^m(\sigma)\right|\propto
\sqrt{(\sigma-\sigma_0)^2+\frac{1}{\tau_0}^2}\,.\label{detMrot}
\end{equation}
Thus, the QNM frequencies are found by evaluating the (complex) matrix
${\bf M}$ for real values of the frequency $\sigma$, finding the
minima of the modulus of its determinant.  The standing wave approach
has thus been generalized to rotating stars.

%%%%%%%%%%%%%%%%%%%%%%%%%%%%%%%%%%%%%%%%%%%%%%
\subsection{Spectral methods for stellar oscillations}\label{spectral}
%%%%%%%%%%%%%%%%%%%%%%%%%%%%%%%%%%%%%%%%%%%%%%
We now summarize the procedure to solve the 2D-perturbed equations
using {\it spectral methods}.  They are, indeed, very powerful to
solve 2D-differential equations, and
particularly useful to implement boundary conditions. For a general
discussion on spectral methods we refer the reader to \cite{spectr}.
%%%%%%%%%%%%%%%%%%%%%%%%%%%%%%%%%%%%%%%%%%%%%%
\subsubsection{Chebyshev polynomials}\label{cheby}
%%%%%%%%%%%%%%%%%%%%%%%%%%%%%%%%%%%%%%%%%%%%%%
We expand all functions of $r$ in Chebyshev polynomials:
\begin{equation}
f(x)=\sum_{n=0}^\infty a_nT_n(x)~~~~~{\rm with}~~~~~
T_n(\cos(u))=\cos(n~u)~~~~n=0,1,\dots
\end{equation}
which satisfy the orthogonality relations
\begin{equation}
\int_{-1}^1T_m(x)T_n(x)\frac{dx}{\sqrt{1-x^2}}
=\frac{\pi}{2}\left(1+\delta_{m0}\right)\delta_{mn}
\label{orthocheb}
\end{equation}
($n,m=0,1,\dots$, $i=1,2,\dots$). The variable $x\in [-1,1]$ is
related to $r\in [a,b]$ by the following equation
\begin{equation}
x=\frac{2r-b-a}{b-a}\in[-1,1]\,.\label{rescale}
\end{equation}
\begin{itemize}
\item{} Integrals on Chebyshev polynomials can be evaluated using the
Gaussian quadrature method \cite{NR}: truncating the polynomial
expansion at $n=K$ we get
\begin{equation}
\int_{-1}^1g(x)\frac{dx}{\sqrt{1-x^2}}=\frac{\pi}{K+1}
\sum_{n=0}^Kg(x_n),
\end{equation}
where the collocation points are
\begin{equation}
x_n=\cos\left(\frac{\pi(n+1/2)}{K+1}\right)~~~~~n=0,1,\dots,K\,.
\end{equation}
\item{} The derivative of a function can be expressed as follows
\begin{equation}
f'(x)=\sum_{n,m}^{0,K}(D_{mn}a_n)T_m\label{dermatrrep}
\end{equation}
with
\begin{eqnarray}
D_{Kn}&=&0\nn\\
D_{k-1\,n}&=&D_{k+1\,n}+2k\delta_{kn}~~~k=2,\dots,K\nn\\
D_{0n}&=&\frac{1}{2}\left[D_{2\,n}+2\delta_{1n}\right]\label{defder}\,.
\end{eqnarray}
\item{}Given a function $V(x)$, and a function $f(x)$ with Chebyshev expansion
\begin{equation}
f(x)=\sum_{n=0}^Ka_nT_n(x)\,,
\end{equation}
the expansion of $V(x)f(x)$ is
\begin{equation}
V(x)f(x)=\sum_{n=0}^Kb_nT_n(x)\,,
\end{equation}
where $b_n=V_{nm}a_m$ with
\begin{equation}
V_{nm}=\frac{2-\delta_{m0}}{K+1}\sum_{k=0}^KV(x_k)
T_n(x_k)T_m(x_k)\,.\label{Vnm}
\end{equation}
\end{itemize}
%%%%%%%%%%%%%%%%%%%%%%%%%%%%%%%%%%%%%%%%%%%%%%%%%%%%%%%%%%%%%%%%%%%%%%%%%%%%%
\subsubsection{Associated Legendre polynomials}\label{legendre}
%%%%%%%%%%%%%%%%%%%%%%%%%%%%%%%%%%%%%%%%%%%%%%%%%%%%%%%%%%%%%%%%%%%%%%%%%%%%%
We expand all functions of $\theta$ in the basis of the associated
Legendre polynomials
\begin{equation}
P^{lm}(y) = (-1)^m(1-y^2)^{m/2}\frac{d^m}{dy^m}P^l(y)\label{defp0},
\end{equation}
with
\begin{equation}
y=\cos\theta\in[-1,1]
\end{equation}
and $m$ fixed.
They are related to scalar spherical harmonics:
\begin{eqnarray}
Y^{lm}(\theta,\phi)&=&\sqrt{\frac{2l+1}{4\pi}\frac{(l-m)!}{(l+m)!}}
P^{lm}(\cos\theta)e^{\ii m\phi}\,~~~~~~{\rm if}~m\ge 0\nn\\
Y^{lm}(\theta,\phi)&=&(-1)^l(Y^{l\,-m})^*~~~~~~~~~~~~~
~~~~~~~~~~~~~~~~~{\rm if}~m<0\,.
\end{eqnarray}
Therefore, expanding a function in circular harmonics $e^{\ii m\phi}$
and in associated Legendre polynomials is equivalent to expand it in
spherical harmonics.  $P^{lm}$'s are eigenfunctions of the Laplacian
operator,
\begin{equation}
\left(\pa_{\theta}^2+\cot\theta\pa_\theta-\frac{m^2}{\sin^2\theta}
\right)P^{lm}=-l(l+1)P^{lm}\,,
\end{equation}
and, assuming for simplicity of notation $m\ge0$, have the 
following asymptotic behaviour near the $z$-axis (see
(\ref{defp0}) ):
\begin{equation}
P^{lm}\sim(1-y^2)^{m/2}=
(\sin\theta)^m~~~~{\rm if}~~\theta\simeq 0,\pi \,.
\label{asymptax}
\end{equation}
This is the asymptotic behaviour of any function of
$\theta$ which is regular on the $z$-axis.  Let us consider a function
$f(r,\theta,\phi)$,  regular in $r=0$ and on the $z$-axis; let us expand it
in circular harmonics
\begin{equation}
f(r,\theta,\phi)=\sum_mf^m(r,\cos\theta)e^{\ii m\phi}\,.
\end{equation}
The regularity of $f(r,\theta,\phi)$  near the $z$-axis implies that
(see for instance \cite{BGM})
\begin{equation}
\lim_{\theta=0,\pi}\frac{f^m(r,\cos\theta)}{(\sin\theta)^m}=\hbox{finite}\,.
\end{equation}
Therefore, the $f^m(r,\cos\theta)$  can be expanded in the polynomials
$\{P^{lm}\}_{l=|m|,\dots}$ with $m$ fixed:
\begin{equation}
f^m(r,y)=\sum_{l=|m|}^\infty a^{lm}(r)P^{lm}(y)\,.\label{fap}
\end{equation}
Thus, associated Legendre polynomials $P^{lm}$ (with $m$
fixed) are a complete basis for all functions of $\theta$ with the
asymptotic behaviour (\ref{asymptax}).

In order to apply the Gaussian quadrature method to associated
Legendre's polynomials, we notice that the polynomials
\begin{equation}
\bar P^{lm}(y)\equiv\frac{P^{lm}(y)}{(1-y^2)^{m/2}}~,
\label{defpbar}
\end{equation}
form, for each $m$, a complete basis, with orthogonality relation
\begin{equation}
\int_{-1}^1\bar P^{lm}(y)\bar P^{l'm}(y)(1-y^2)^mdy=
\frac{2}{2l+1}\frac{(l+m)!}{(l-m)!}\delta_{ll'}\,.\label{orthopbar}
\end{equation}
The $\bar P^{lm}(y)$'s  are a particular case of Jacobi's polynomials 
$J_l^{(\alpha,\beta)}(y)$ defined by \cite{NR}
\begin{equation}
\int_{-1}^1 J_l(y)J_{l'}(y)(1-y)^\alpha(1+y)^\beta\propto  \delta_{ll'}\,,
\end{equation}
where $\alpha=\beta=m$. Therefore, Gaussian integration takes the form
\begin{equation}
\int_{-1}^1f^m(y)P^{lm}(y)dy=\sum_kw_k\frac{f(y_k)P^{lm}(y_k)}{1-y_k^2},
\end{equation}
where $y_k$ and $w_k$ are the collocation points and weights for the
Jacobi polynomials with $\alpha=\beta=m$. In particular, the
coefficients of the expansion (\ref{fap}) are
\begin{equation}
a^{lm}=\frac{2l+1}{2}\frac{(l-m)!}{(l+m)!}
\sum_kw_k\frac{f(y_k)P^{lm}(y_k)}{1-y_k^2}\,.
\end{equation}
%%%%%%%%%%%%%%%%%%%%%%%%%%%%%%%%%%%%%%%%%%%%%%%%%%%%%%%%%%%%%%%%%%%%%%%%%%%%%
\subsubsection{Differential equations and boundary conditions}
%%%%%%%%%%%%%%%%%%%%%%%%%%%%%%%%%%%%%%%%%%%%%%%%%%%%%%%%%%%%%%%%%%%%%%%%%%%%%
Let us consider a one-dimensional, first order, differential equation
\begin{equation}
Z'(x)+V(x)Z(x)=0\label{zpvz}
\end{equation}
with $x\in[-1,1]$. If we expand $Z(x)$ in the basis of Chebyshev
polynomials, $T_n(x)$, truncating the expansion at $n=K$, i.e.
\begin{equation}
Z(x)=\sum_{n=0}^Ka_nT_n(x),
\end{equation}
the differential equation (\ref{zpvz}) becomes  an algebraic
equation:
\begin{equation}
\sum_{k=0}^K(D_{nk}+V_{nk})a_k=0~~~~~(n=0,\dots,K)\,,\label{dva}
\end{equation}
where $D_{nk}$ and $V_{nk}$ are defined in (\ref{defder}), (\ref{Vnm}).

The boundary conditions needed to solve Equation (\ref{zpvz}) can be
implemented using the so-called $\tau$-{\it method}: we cut the last
row of (\ref{dva}), i.e. we set
\begin{equation}
\sum_{k=0}^K(D_{nk}+V_{nk})a_k=0~~~~~(n=0,\dots,K-1),
\label{dva1}
\end{equation}
and replace the row with the boundary condition; for instance, if
we know that $Z(x_0)=z_0$, the last row of the matrix equation will be
replaced with
\begin{equation}
\sum_{k=0}^KT_k(x_0)a_k=z_0\,.\label{bc1}
\end{equation}
The differential equation (\ref{zpvz}), with the boundary condition
$Z(x_0)=z_0$, thus reduces to a matrix equation which can be solved by
$LU$ decomposition \cite{NR}.

This approach can easily be generalized to higher order differential
equations (by replacing more rows for the associated boundary
conditions), to systems of coupled differential equations, and to
partial differential equations in $r,\theta$. In this case, each
function is expanded in the basis $\{T_n(x),P^{lm}(y)\}_{n,l}$ as
follows
\begin{equation}
f^m(x,y)=\sum_{n=0}^K\sum_{l=|m|}^La^{lm}_nT_n(x)P^{lm}(y),
\label{doubleexp}
\end{equation}
where sums over Chebyshev's and associated Legendre's polynomials are
truncated to the orders $K$ and $L$, respectively; the coefficients
$a^{lm}_n$ can be organized as the components of a vector, defining
the collective index
\begin{equation}
i(l,n)=\left[l-|m|\right](K+1)+n+1=1,2,\dots,\left[L-|m|+1\right](K+1)
\end{equation}
and setting
\begin{equation}
\hat a^m_i=\hat a^m_{i(l,n)}\equiv a^{lm}_n\,.
\end{equation}
In terms of the expansion (\ref{doubleexp}), PDEs in $r,\theta$ (in
our case, the equations which describe the perturbations of rotating
stars) transform into an algebraic equation.  As an example of the double
expansion (\ref{doubleexp}), in Appendix \ref{RWspher2d} we show how
to solve the Regge-Wheeler equation  as a partial differential equation
 in $r,\theta$.
%%%%%%%%%%%%%%%%%%%%%%%%%%%%%%%%%%%%%%%%%%%%%%%%%%%%%%%%%%%%%%%
\section{A test of the method: oscillations of slowly rotating stars}
\label{slowly}
%%%%%%%%%%%%%%%%%%%%%%%%%%%%%%%%%%%%%%%%%%%%%%%%%%%%%%%%%%%%%%%
As a test of our method, we have solved the equations which describe
the perturbations of a slowly rotating stars.
Following \cite{Hartle}, in this case the metric and fluid velocity of
the unperturbed star are
\begin{eqnarray}
(ds^2)^{(0)}&=&g^{(0)}_{\mu\nu}dx^\mu dx^\nu=e^{-\nu(r)}dt^2+e^{\lambda(r)}dr^2
+r^2(d\theta^2+\sin^2\theta d\phi^2)-2\omega(r)r^2\sin^2\theta dtd\phi\nn\\
u^{(0)\,\mu}&=&(e^{-\nu/2},0,0,\Omega e^{-\nu/2})\label{slowstar}
\end{eqnarray}
where $\omega(r)$ describes the dragging of the inertial frames,
and all quantities are expanded at first order in the
angular velocity of the star, $\Omega$.  We introduce a rotation
parameter $\epsilon$ defined by
\begin{equation}
\epsilon = \Omega/\sqrt{M/R^3}\,.
\end{equation}
In the metric (\ref{slowstar}) spherical symmetry is broken only by
the term $\omega$; when Einstein's equations are perturbed about
(\ref{slowstar}), only terms which are linear in $\omega$ are
retained, i.e. we keep terms up to order $O(\epsilon)$.  As a
consequence, perturbations with index $l$ are coupled with
perturbations with indexes $l\pm1$ through terms that are of order
$O(\epsilon)$, the analytical form of which can explicitly be
derived.  Therefore, when we transform the perturbed equations using
the double spectral decomposition described in Section \ref{spectral},
the resulting algebraic equations have a particularly simple form: the
relevant matrix is ``almost-block-diagonal'', each block corresponding
to one value of $l$; the off-diagonal blocks couple $l\leftrightarrow
l\pm1$.

These equations can also be obtained in a different way, i.e.  by
expanding in Chebyshev polynomials
the system of ordinary differential equations in $r$ derived
by Kojima in \cite{Kojima1992} (hereafter, K1).
This follows from the fact that Kojima's equations are
derived by expanding the perturbed Einstein's equations in spherical
harmonics.

The general structure of  Kojima's equations is the following:
\begin{eqnarray}
{\cal L}^{pol}[H_0^{lm},K^{lm};\sigma]
&=&m{\cal E}[H_0^{lm},K^{lm};\sigma]
+{\cal F}^{(\pm)}[Z_{RW}^{l\pm1\,m};\sigma]\nonumber\\
{\cal L}^{ax}[Z_{RW}^{lm};\sigma]&=&
m{\cal N}[Z_{RW}^{lm};\sigma]+
{\cal D}^{(\pm)}[H_0^{l\pm1\,m},K^{l\pm1\,m};\sigma].
\label{kojstruct}
\end{eqnarray}
Here ${\cal L}^{pol}$, ${\cal L}^{ax}$ are operators of order
$O(\epsilon^0)$, which describe the perturbations of the star in the
non rotating case; ${\cal E}$, ${\cal F}^{(\pm)}$, ${\cal N}$, ${\cal
D}^{(\pm)}$ are $O(\epsilon^1)$ operators, which provide the
corrections due to rotation.  These equations have been integrated
numerically in \cite{Kojima1993} (see also \cite{KojimaPTP}) using a
very strong simplification: the couplings $l\leftrightarrow l\pm1$
were neglected (i.e. ${\cal F}^{(\pm)}$ and ${\cal D}^{(\pm)}$ were
set to zero).

Moreover, the equations were solved iteratively, finding the solution
for $\epsilon=0$ first, and then replacing it in the terms ${\cal
E}[H_0^{lm},K^{lm};\sigma]$ and ${\cal N}[Z_{RW}^{lm};\sigma]$ of
eqs. (\ref{kojstruct}). In this way, the right-hand sides of
eqs. (\ref{kojstruct}) become source terms.

We stress that with our approach we do not need these simplifications
anymore; in particular, we do not need to neglect the couplings,
because we can handle the mixing among perturbations with different
$l$'s using the spectral methods and the generalized standing wave
approach.

It should also be mentioned that, when coupling terms are included in
the perturbed equations, in order to have a good numerical behaviour
of the perturbations near the center of the star we need to use a set
of variables (in particular the $O(\epsilon^1)$ terms) different from
that used in \cite{KojimaPTP}.  The equations for the
new variables are given explicitly
in Appendix \ref{couplkoj}.

%%%%%%%%%%%%%%%%%%%%%%%%%%%%%%%%%%%%%%%%%%%%%%%%%%%%%%%%%%%%%%%
\subsection{Comparison with existing results}
%%%%%%%%%%%%%%%%%%%%%%%%%%%%%%%%%%%%%%%%%%%%%%%%%%%%%%%%%%%%%%%
As mentioned above, in \cite{Kojima1993} (hereafter K2) the equations
of stellar perturbations have been integrated for a slowly rotating
star, neglecting $l\leftrightarrow l\pm1$ couplings, and the QNM
frequencies have been found; to reproduce these results we have used the
same set of variables as in K1,  the same equation of state (EOS)
i.e.  the polytropic EOS $p=K\rho^2$, and  we have computed the fundamental mode
($f$-mode) frequency, $\sigma_f$.

\begin{figure}[ht]
\includegraphics[width=6cm,angle=270]{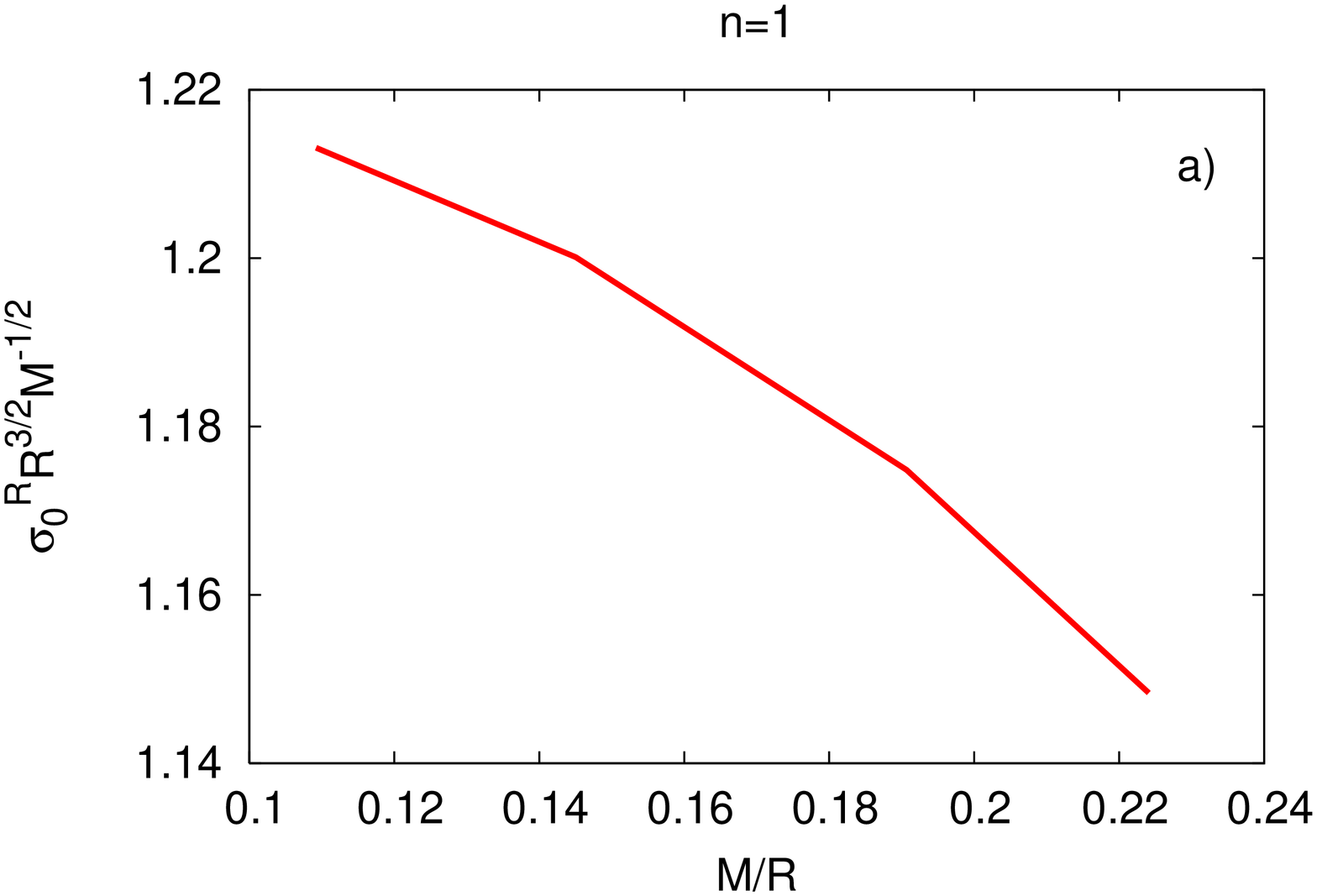}
\includegraphics[width=6cm,angle=270]{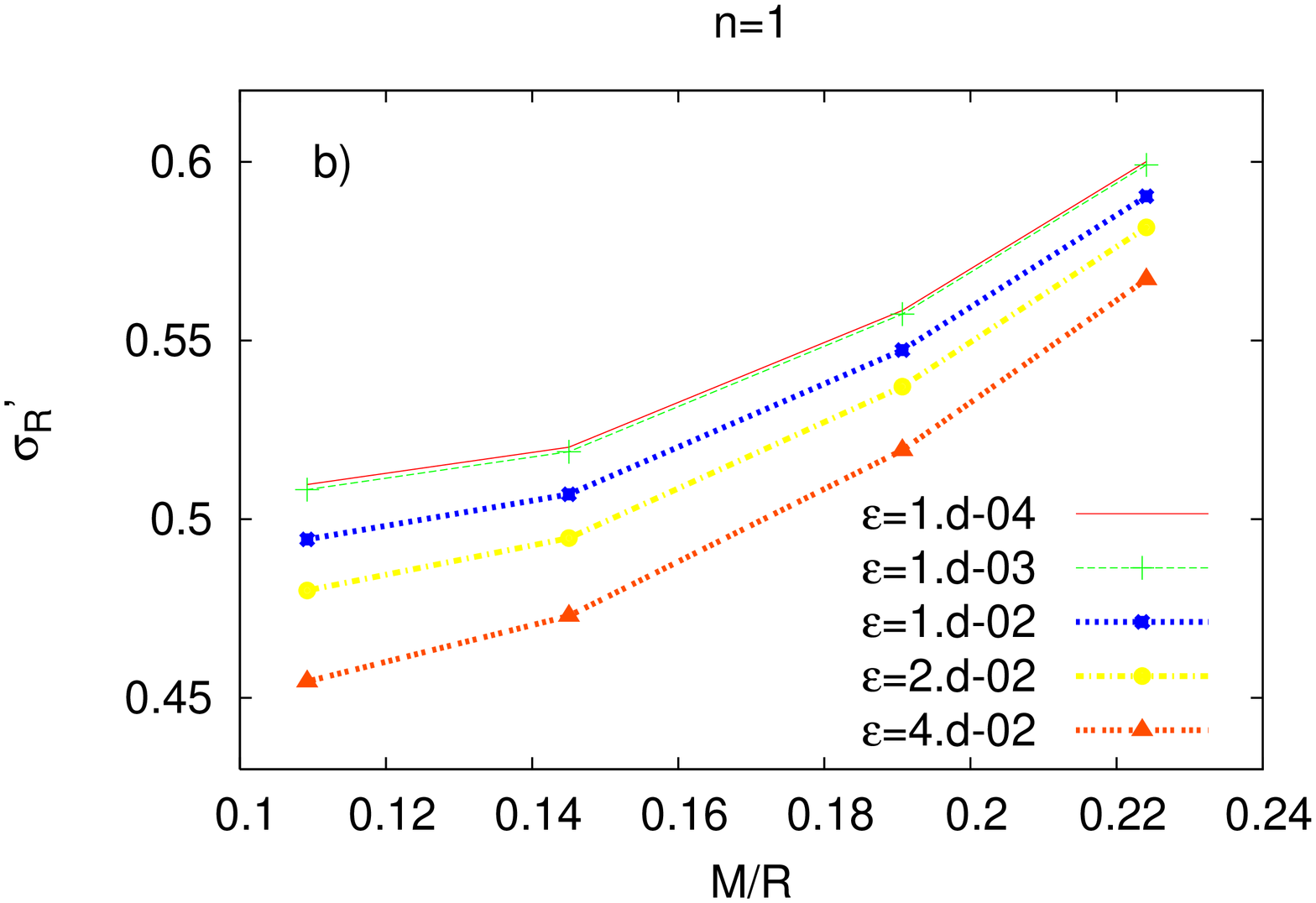}
\caption{The real part of the $f$-mode frequency of a non rotating,
polytropic star with $n=1$, is plotted as a function of the stellar
compactness $M/R$ (a); the frequency shift due to rotation,
$\sigma_R'$ is plotted versus compactness for different values of
$\epsilon$ (b).  As in K2, couplings among different $l$'s are
neglected.
\label{koj93nu}}
\end{figure}

In K2 the real and imaginary parts of $\sigma_f$ are fitted as functions of the
rotation parameter $\epsilon$  as follows:
\begin{eqnarray}
\sigma_f^R&=&\sigma_0^R(1+m\epsilon\sigma_R') + O(\epsilon^2)\,,\nonumber\\
\sigma_f^I&=&\sigma_0^I(1+m\epsilon\sigma_I') + O(\epsilon^2)\,.\label{fitdecoupl}
\end{eqnarray}
The value of $\sigma_0^R$ we find, properly normalized, is plotted
versus the stellar compactness, $M/R$, in Figure \ref{koj93nu} a). The
values are in excellent agreement with the results shown in Figure 1
of K2, for $n=1$.

The correction due to rotation, $\sigma_R'$, is
plotted versus $M/R$ in Figure \ref{koj93nu} b) for different values
of $\epsilon$.  We find that for $\epsilon \lesssim 10^{-3}$ the
corresponding curves are indistinguishable, and coincide with the
$n=1$ curve shown in Figure 1 of K2.  However, for $\epsilon\gtrsim
10^{-3}$ different $\epsilon$ correspond to different curves, and
the fit (\ref{fitdecoupl}) becomes inaccurate:  $\sigma_R'$ is no longer a
constant, and further corrections to (\ref{fitdecoupl}) are of order
$O(\epsilon^2)$, as expected theoretically.
%%%%%%%%%%%%%%%%%%%%%%%%%%%%%%%%%%%%%%%%%%%%%%%%%%%%%%%%%%%%%%%
%%%%%%%%%%%%%%%%%%%%%%%%%%%%%%%%%%%%%%%%%%%%%%%%%%%%%%%%%%%%%%%

%%%%%%%%%%%%%%%%%%%%%%%%%%%%%%%%%%%%%%%%%%%%%%%%%%%%%%%%%%%%%%%
\begin{figure}[ht]
\includegraphics[width=6cm,angle=270]{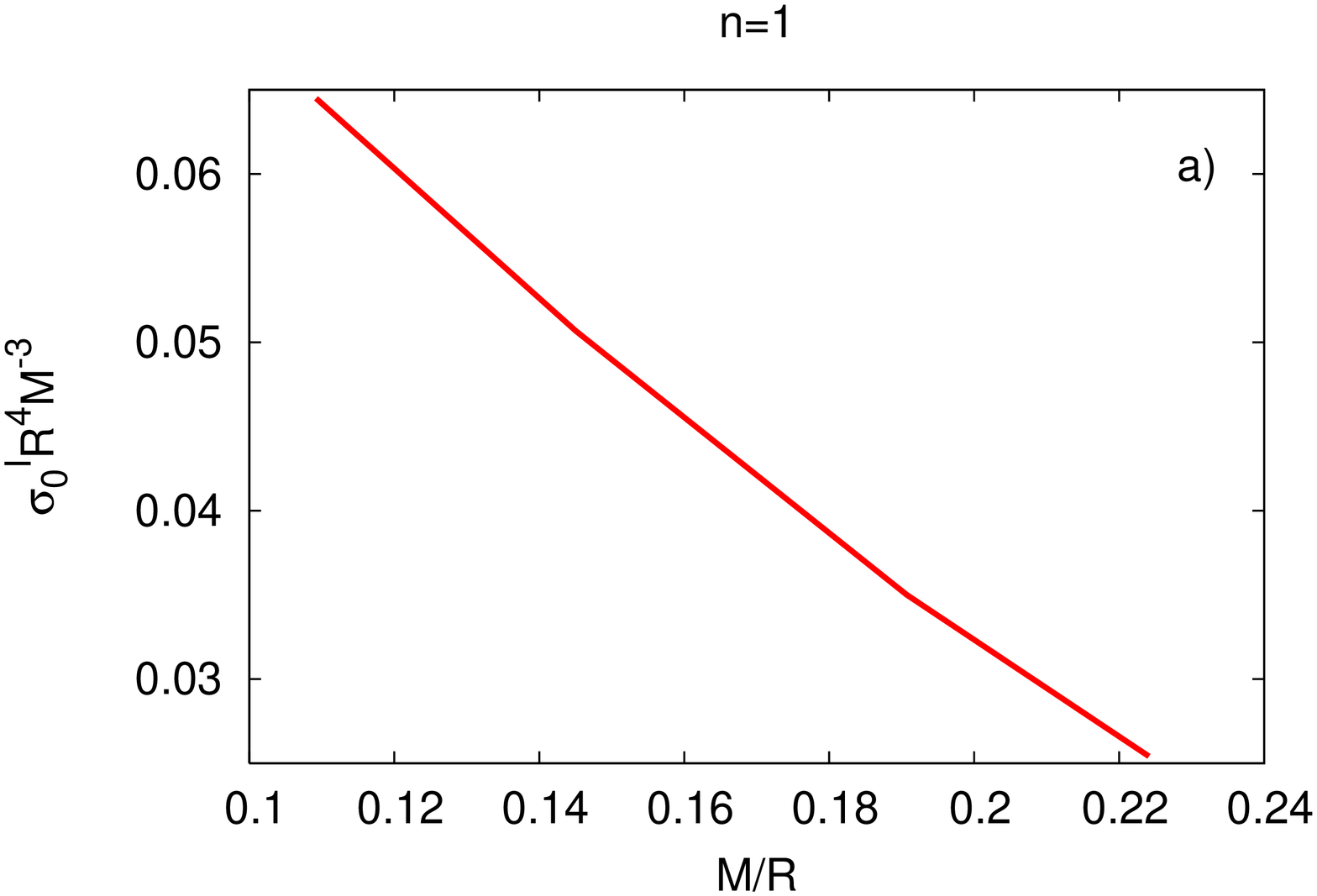}
\includegraphics[width=6cm,angle=270]{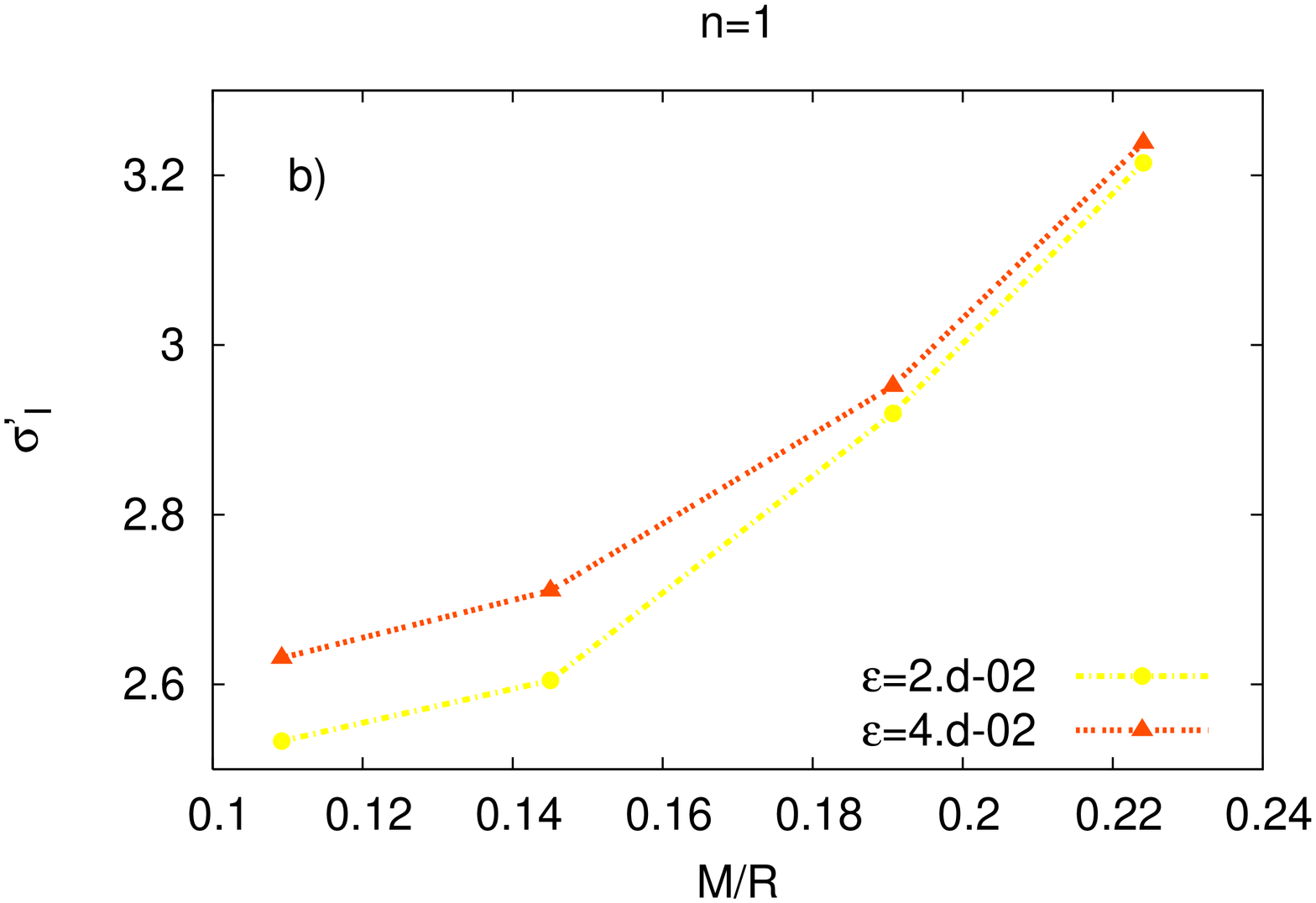}
\caption{The imaginary part of the $f$-mode frequency (a) and the
corrections due to rotation (b) are plotted as in Figure
\ref{koj93nu}.  }
\label{koj93damp}
\end{figure}
%%%%%%%%%%%%%%%%%%%%%%%%%%%%%%%%%%%%%%%%%%%%%%%%%%%%%%%%%%%%%%%

In Figure \ref{koj93damp} we plot the imaginary part of the $f$-mode
frequency, $\sigma_0^I$, and the corresponding rotational correction,
$\sigma_I'$, as in Figure \ref{koj93nu}.  $\sigma_I'$ is plotted only
for $\epsilon\ge 10^{-2}$, because for smaller values our numerical
approach becomes inaccurate. They agree with the values given in
Figure 2 of K2, with differences of order $O(\epsilon^2)$.
%%%%%%%%%%%%%%%%%%%%%%%%%%%%%%%%%%%%%%%%%%%%%%%%%%%%%%%%%%%%%%%
\subsection{Including the couplings}
%%%%%%%%%%%%%%%%%%%%%%%%%%%%%%%%%%%%%%%%%%%%%%%%%%%%%%%%%%%%%%%
We now apply our method to solve eqs. (\ref{kojstruct}) in full, i.e.
including couplings among different $l$'s.  The numerical
implementation of the equations presents a problem: there are terms in
the equations which depend on $\left(\frac{\partial
p}{\partial\rho}\right)^2$ and on $\left(\frac{\partial^2
p}{\partial\rho^2}\right)$, which strongly diverge on the stellar
surface. This divergence is particularly problematic when spectral
methods are used, but it can be cured through a regularization
procedure \cite{BGM}. Such regularization goes beyond the scope of the
present paper, where we only want to discuss a simple implementation
of our approach.  Therefore, we solve the perturbed equations in the
case of a constant density, slowly rotating star, such that the
divergent terms vanish.

We consider two background models: $A$, with $M/R=0.2$ and $B$, with
$M/R=0.1$. Mass and radius for  assigned values of the
central density are given in Table \ref{TabA}.
%%%%%%%%%%%%%%%%%%%%%%%%%%%%%%%%%%%%%%%%%%%%%%%%%%%%%%%%%%%%%%%
\begin{table}[htbp]
\begin{center}
\begin{tabular}{|c|c|c|c|c|}
\hline
&$\rho$ (g/cm$^3$) & $M/M_\odot$ & $R$(Km) & $M/R$ \\
\hline
$A$ & $10^{15}$ & 1.11 & 8.08 & 0.2 \\\hline
$B$ & $10^{15}$ & 0.40 & 5.75 & 0.1  \\\hline
\end{tabular}
\caption{Parameters of the constant density stellar models $A$ and $B$
we use as a background.
\label{TabA}}
\end{center}
\end{table}
%%%%%%%%%%%%%%%%%%%%%%%%%%%%%%%%%%%%%%%%%%%%%%%%%%%%%%%%%%%%%%%
The explicit form of the equations and the boundary conditions in
$r=0$ and $r=R$ are discussed in Appendix \ref{couplkoj}.  As
mentioned above, when couplings are included the equations derived in
K1 are very unstable when integrated near the center.  For this reason
we introduce a new set of variables, which satisfy a new set of
equations shown in Appendix \ref{couplkoj}; the appropriate boundary
conditions in $r=0$ and $r=R$ are also shown.  Once we assign the
value of the harmonic index $m$, these equations couple polar and
axial perturbations with $|m|\le l\le L$.

The equations have been integrated for $m = \pm 2$.  We do not set
$|m|<2$ because in that case dipolar ($l=1$) perturbations have to be
taken into account, which are described by equations different from
those we consider in this paper.  We would like to stress that
rotational corrections to mode frequencies with $m\neq0$ are much
larger than those to mode frequencies with $m=0$.  In K2, Kojima
suggested that, at lowest order in $\epsilon$, the frequency shift is
proportional to $m$.  This is consistent with the results of our
numerical integration: the relative frequency shifts found in
\cite{DSF}, where $m=0$ perturbations were studied in full general
relativity, are an order of magnitude smaller than the relative shifts
we find for $m=\pm 2$.

If the star does not rotate, for any assigned value of $l$ there is a
corresponding $f$-mode frequency, which is the same for all $m$'s.  If
the star rotates, due to the couplings the $f$-mode belonging to an
assigned $l$ acquires contributions from different $l$'s, and its
frequency and damping time change.  Furthermore, the degeneracy in $m$
is broken by rotation.

\begin{figure}[htbp]
\includegraphics[width=6cm,angle=270]{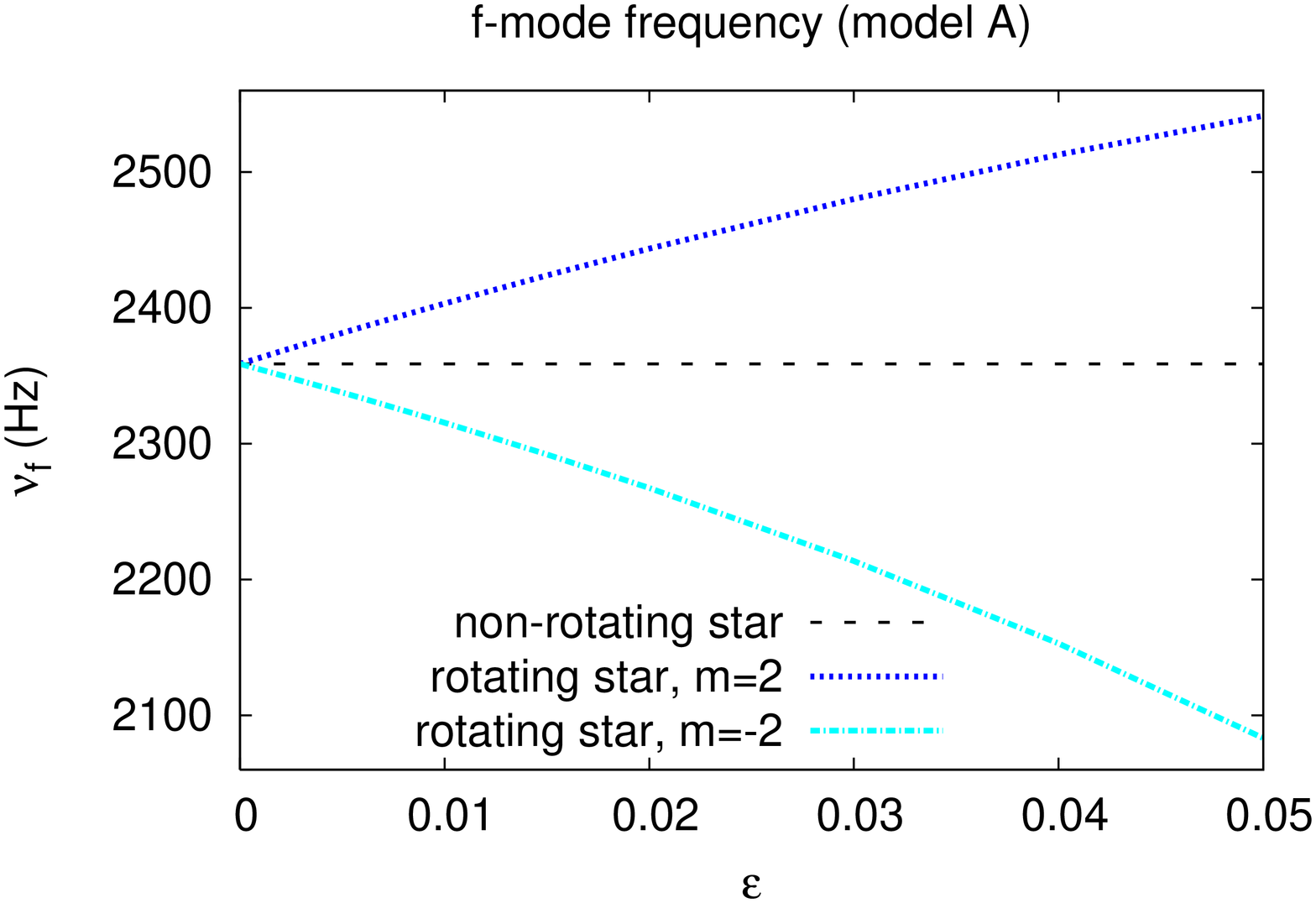}
\includegraphics[width=6cm,angle=270]{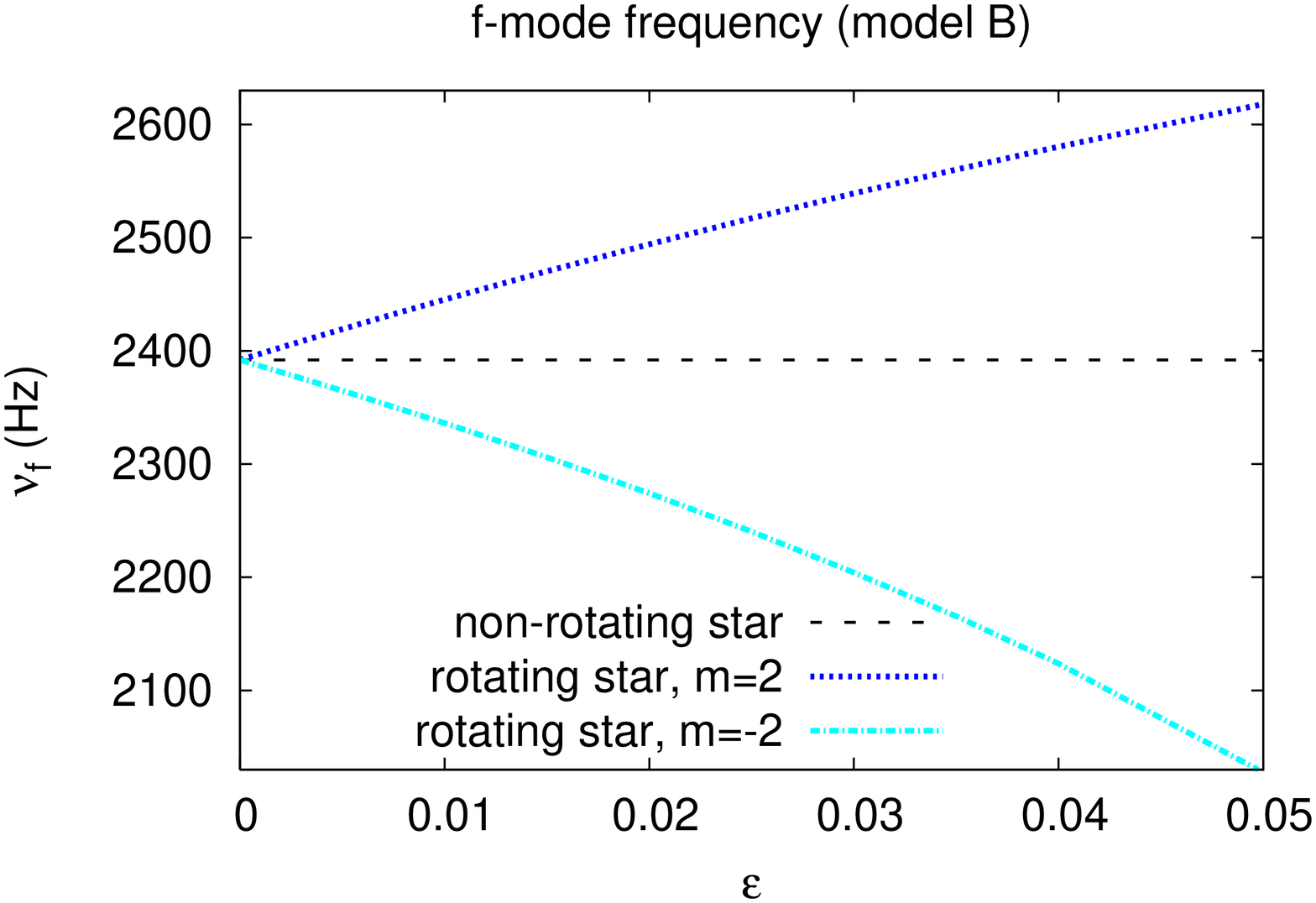}
\caption{The real part of the $f$-mode frequency,
$\nu_f=\sigma^R_f/(2\pi)$, is plotted versus the rotation parameter
$\epsilon$. The data refer to two models of constant density star (see
text).  We include couplings up to $l=4$.  The dashed line refers
to the non rotating star; the dotted lines  refer to the modes $m=\pm
2$ of the slowly rotating star.
\label{fmode}
}
\end{figure}

The real part of the $f$-mode frequency, $\nu_f=\sigma^R_f/(2\pi)$, is
plotted as a function of the rotation parameter in Figure \ref{fmode}
for the two considered stellar models. The solid line represents the
frequency of the $l=2$ $f$-mode of the non-rotating star. The dashed
lines are the frequencies of the lowest lying fundamental mode of the
rotating star, with $m=2$ and $m=-2$, assuming $L=4$.  Our
calculations refer to $\epsilon\le0.05$, since for higher values the
slow rotation approximation becomes inaccurate and the results cannot
be trusted anymore.
%%%%%%%%%%%%%%%%%%%%%%%%%%%%%%%%%%%%%%%%%%%%%%%%%%%%%%%%%%%%%%%
%%%%%%%%%%%%%%%%%%%%%%%%%%%%%%%%%%%%%%%%%%%%%%%%%%%%%%%%%%%%%%%

\begin{figure}[htbp]
\includegraphics[width=6cm,angle=270]{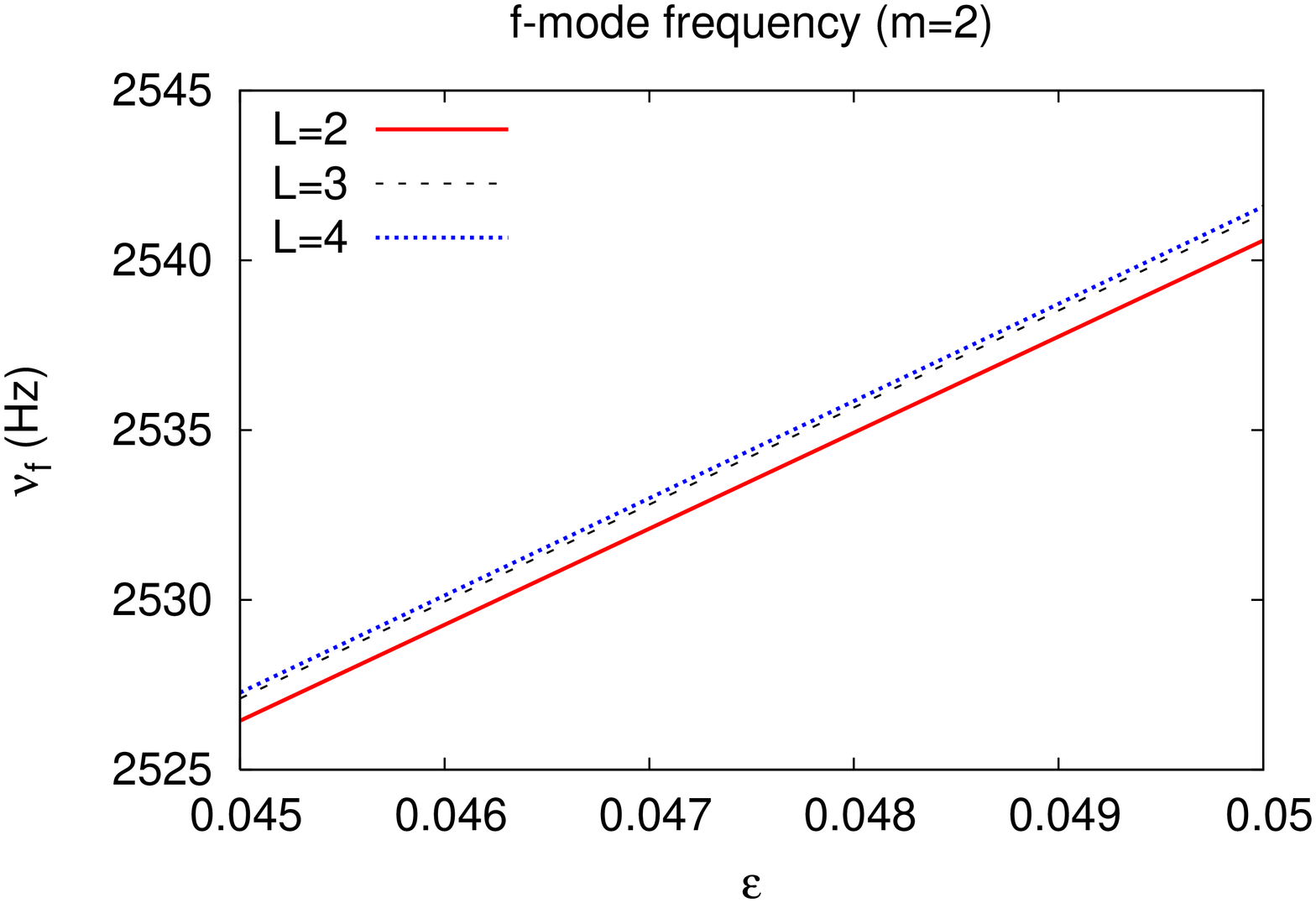}
\includegraphics[width=6cm,angle=270]{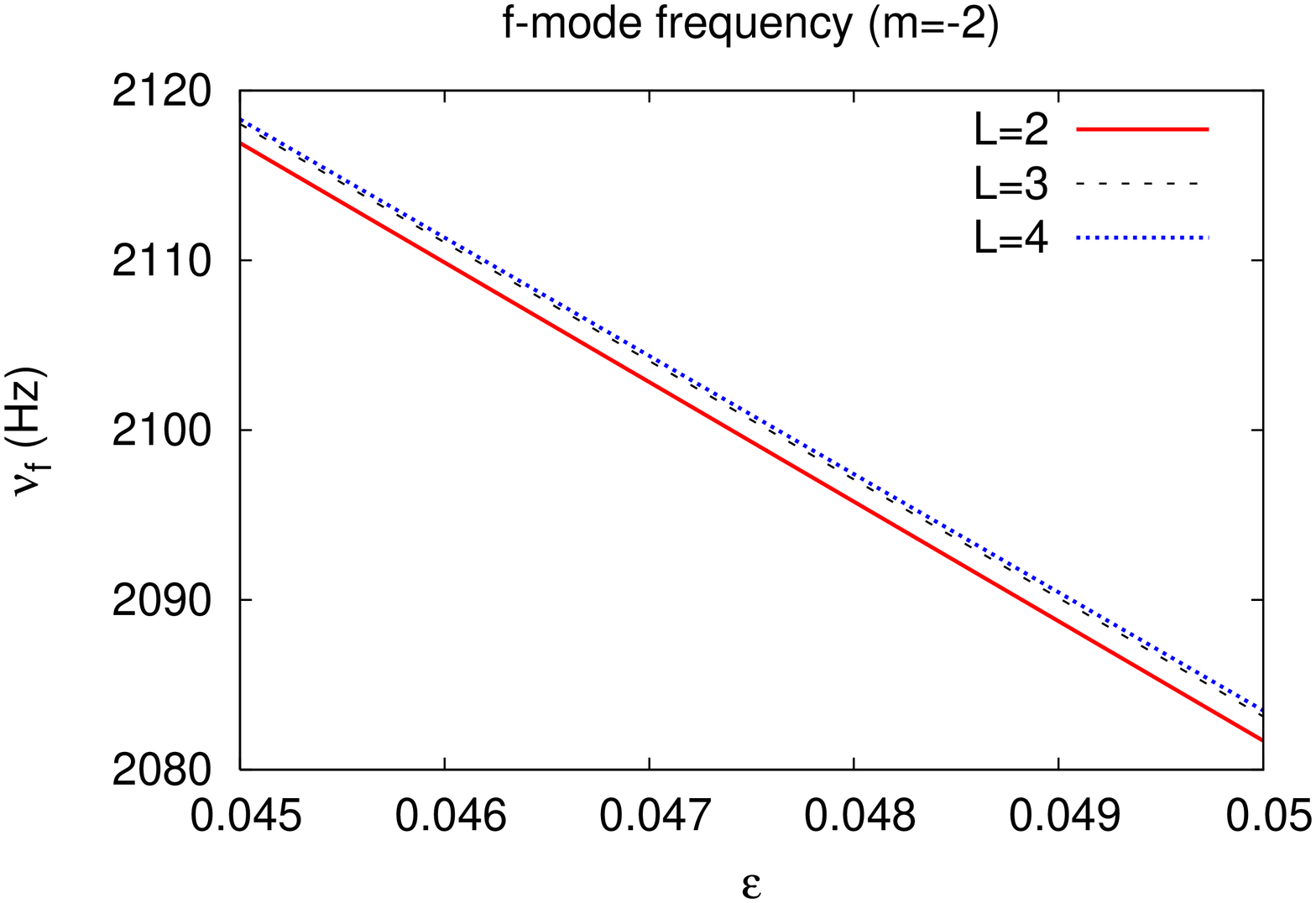}
\caption{Details of left panel of Figure \ref{fmode}: we show 
the contribution of the different couplings to the $f$-mode
frequency, for model $A$. The different curves are obtained by
including in the perturbed equations couplings up to $l=L$.
\label{detail}}
\end{figure}

Figure \ref{detail} shows, in a smaller range of the rotation
parameter, the frequency $\nu_f$ computed by truncating the expansion in
$l$ at $L=2$ (i.e. without couplings), $L=3$ and $L=4$. It is evident
that for slowly rotating stars the contribution of the couplings is a
small correction, and that there is convergence as $L$ grows.
%%%%%%%%%%%%%%%%%%%%%%%%%%%%%%%%%%%%%%%%%%%%%%%%%%%%%%%%%%%%%%%
%%%%%%%%%%%%%%%%%%%%%%%%%%%%%%%%%%%%%%%%%%%%%%%%%%%%%%%%%%%%%%%

\begin{figure}[htbp]
\includegraphics[width=6cm,angle=270]{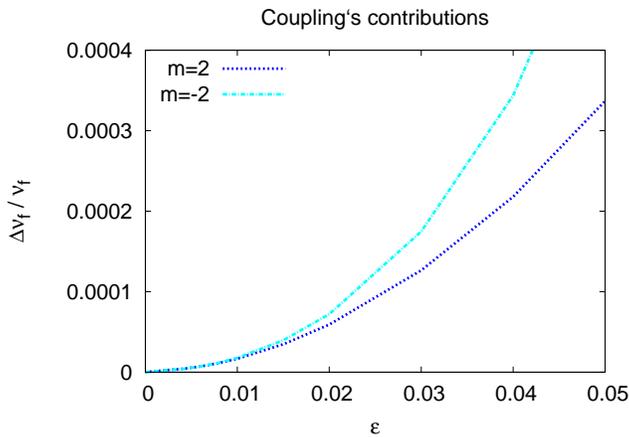}
\caption{Frequency shift as given by equation
(\ref{Deltaf}) for model A.\label{couplings}}
\end{figure}

The relative frequency shift due to the couplings is well approximated
by a quadratic behaviour in $\epsilon$:
\begin{equation}
\frac{\Delta\nu_f}{\nu_f}=\frac{\nu_f^{L=4}(\epsilon)
  -\nu_f^{L=2}(\epsilon)}{\nu_f^{\epsilon=0}}
\simeq\sigma''\epsilon^2\, ,\label{Deltaf}
\end{equation}
as shown in Figure \ref{couplings}.
%%%%%%%%%%%%%%%%%%%%%%%%%%%%%%%%%%%%%%%%%%%%%%%%%%%%%%%%%%%%%%%
%%%%%%%%%%%%%%%%%%%%%%%%%%%%%%%%%%%%%%%%%%%%%%%%%%%%%%%%%%%%%%%
Therefore, the contribution of the couplings is of order
$O(\epsilon^2)$, as argued by Kojima in K2, and $\sigma_f^R$ is well
described by a quadratic fit of the form
\begin{equation}
\sigma_f^R=\sigma_0^R(1+m\epsilon\sigma'_R+\epsilon^2\sigma''_R)\,
,\label{fitcoupl}
\end{equation}
where we should remind that terms of order $O(\epsilon^2)$ are of the
same order of the terms which we are neglecting in the perturbed
equations ab initio.
%%%%%%%%%%%%%%%%%%%%%%%%%%%%%%%%%%%%%%%%%%%%%%%%%%%%%%%%%%%%%%%
\begin{figure}[htbp]
\includegraphics[width=6cm,angle=270]{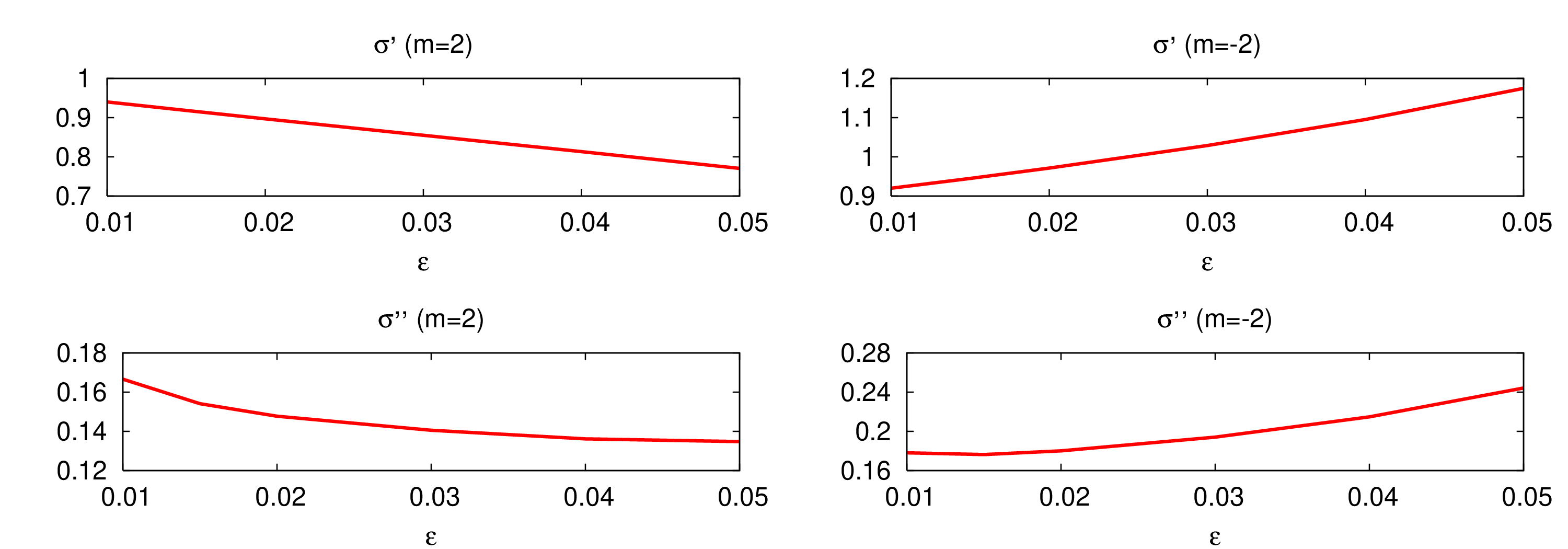}
\caption{The coefficients $\sigma',\sigma''$, plotted as functions of
the rotation rate for model $A$.\label{coefficients}}
\end{figure}
%%%%%%%%%%%%%%%%%%%%%%%%%%%%%%%%%%%%%%%%%%%%%%%%%%%%%%%%%%%%%%%
This fit is accurate up to $\epsilon\sim 10^{-3}$; for larger values
$\sigma'$ and $\sigma''$ are no longer constant, as shown in Figure
\ref{coefficients}: as $\epsilon$ grows, $\sigma'$ changes linearly,
and the change is negative if $m>0$, positive if $m<0$, yielding in
both cases a shift to lower values of the total frequency
$\sigma$. This explains the small asymmetry between $\nu_f(m=2)$ and
$\nu_f(m=-2)$ shown in Figure \ref{fmode}. Notice that deviations from
the fit (\ref{fitcoupl}) are always of order $O(\epsilon^2)$,
consistently with our approximation scheme.  For $0.01<\epsilon<0.05$,
the coefficients $\sigma''$ are $\sim 0.15$ (if $m=2$) and $\sim 0.20$
(if $m=-2$); for $\epsilon<0.01$ they are too small to be correctly
extrapolated with our codes.  Finally, the damping time of the
$f$-mode is shown in Figure \ref{tau} as a function of the rotation
rate.
%%%%%%%%%%%%%%%%%%%%%%%%%%%%%%%%%%%%%%%%%%%%%%%%%%%%%%%%%%%%%%%
\begin{figure}[htbp]
\includegraphics[width=6cm,angle=270]{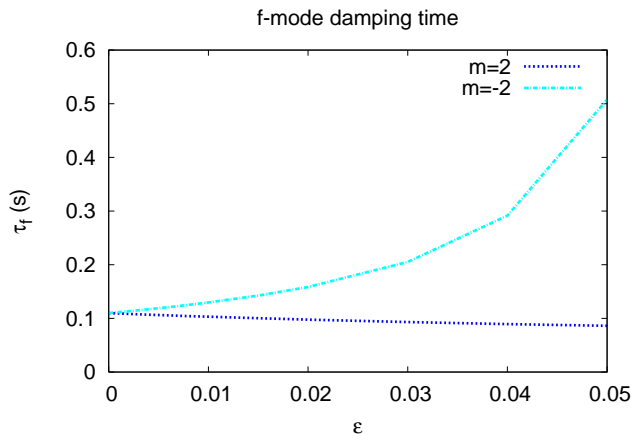}\\
\caption{Damping time of the $f$-mode as a function of the rotation
rate, for model $A$.\label{tau}}
\end{figure}
%%%%%%%%%%%%%%%%%%%%%%%%%%%%%%%%%%%%%%%%%%%%%%%%%%%%%%%%%%%%%%%

It is worth stressing that, as the stellar rotation increases, the
frequency of the counterrotating (i.e. $m=-2$) mode decreases faster
than expected by the simple linear fit (\ref{fitdecoupl}) .
Furthermore, Figure \ref{tau} shows that the damping time of the
counterrotating mode increases sharply, even for small rotation rates.
This indicates that the CFS instability may occur for values of the
rotation rate lower than expected by simple linear estimates.

It should be mentioned that the equations derived for the
perturbations of a slowly rotating star in \cite{Kojima1992} and in
Appendix \ref{couplkoj}, are not appropriate to study the $r$-modes,
because the frequency $\sigma$, which is a dimensionful scale of the
problem, is of the same order as the ``small'' parameter $\epsilon$.
The shift of the $r$-mode frequency due to slow rotation in a
relativistic star has been studied in \cite{LFA}, taking into account
the couplings between perturbations with different $l$'s.
%%%%%%%%%%%%%%%%%%%%%%%%%%%%%%%%%%%%%%%%%%%%%%%%%%%%%%%%%%%%%%%%%%%%%%%%%%%%
\section{Concluding remarks}\label{conclusions}
%%%%%%%%%%%%%%%%%%%%%%%%%%%%%%%%%%%%%%%%%%%%%%%%%%%%%%%%%%%%%%%%%%%%%%%%%%%%
In this paper we propose a new approach to find the quasi-normal mode
frequencies of rotating relativistic stars. We describe the main
features of the general method, and test it in the particular case of
slowly rotating stars. The application of our method to rapidly
rotating stars will be discussed in a forthcoming paper.

We give explicit formulae (whose numerical implementation is
straightforward) which allow to transform functions of $r,\theta$ in
vectors, and systems of coupled differential equations (involving
derivatives in $r,\theta$) in algebraic, matrix equations.
Furthermore, we show that, once the 2D-equations describing stellar
perturbations have been solved, the frequencies and damping times of
slowly damped, quasi-normal modes can be found by looking for the
minima of the determinant of a properly defined matrix, evaluated as
function of real frequency, thus generalizing the standing wave
approach \cite{SW,SW2} to rotating stars.

We have tested the method in the case of slow rotation; the system of
partial differential equations from which we start is formally the
same as in \cite{Kojima1992} (apart from a redefinition of some
variables described in Appendix \ref{couplkoj}).  However, in our
approach we transform that differential system in a system of
algebraic, coupled equations; thus, the advantage of our method is
that it is much easier to handle the couplings among different values
of $l$, which in \cite{Kojima1992} correspond to couplings between
different partial differential equations. 
For this reason we are able to study the shift of the
fundamental mode due to rotation, taking into account the
$l\leftrightarrow l\pm1$ couplings, to our knowledge for the first
time in the literature.

In this paper we show that, as the rotation parameter $\epsilon$ increases, the
frequency of counterrotating modes decreases at a rate higher than
linear in $\epsilon$.  Furthermore, the corresponding damping time
sharply increases, even for small rotation rates.  This suggests that
the CFS instability for a generic mode should occur for values of the rotation rate lower
than expected by simple linear estimates.  
This result complements what found in \cite{SF},
where the equations of stellar perturbations where integrated
in full general relativity looking for neutral modes, and it was shown that
the CFS instability sets in for smaller rotation rates than in Newtonian gravity.

It should be mentioned that an alternative approach to find the QNM
frequencies, based on a characteristic formulation
of the perturbed equations and a complexification of the radial coordinate,
which could be generalized to rotating stars, has recently been proposed \cite{nils}.
%%%%%%%%%%%%%%%%%%%%%%%%%%%%%%%%%%%%%%%%%%%%%%%%%%%%%%%%%%%%%%%%%%%%%%%%%%%%
\begin{acknowledgments} 
We are grateful to Kostas Kokkotas for suggesting us to generalize the
standing wave approach to rotating stars.  We thank Loic Villain for
useful suggestions about the spectral methods, and Jos\`e Pons for
useful discussions.  S.M. is supported by a ``Virgo EGO Scientific
Forum (VESF)'' grant.
\end{acknowledgments}
\appendix
%%%%%%%%%%%%%%%%%%%%%%%%%%%%%%%%%%%%%%%%%%%%%%%%%%%%%%%%%%%%%%%%%%%%%%%%%%%%
\section{Gauge choice}\label{genRW}
%%%%%%%%%%%%%%%%%%%%%%%%%%%%%%%%%%%%%%%%%%%%%%%%%%%%%%%%%%%%%%%%%%%%%%%%%%%%
In this Appendix we denote with $A,B,\dots$ the coordinates $t,r$, and
with $a,b,\dots$ the coordinates $\theta,\phi$; we define
$\gamma_{ab}\equiv{\rm diag}(1,\sin^2\theta)$.
%%%%%%%%%%%%%%%%%%%%%%%%%%%%%%%%%%%%%%%%%%%%%%%%%%%%%%%%%%%%%%%%%%%%%%%%%%%%
\subsection{The generalized Regge-Wheeler gauge}
%%%%%%%%%%%%%%%%%%%%%%%%%%%%%%%%%%%%%%%%%%%%%%%%%%%%%%%%%%%%%%%%%%%%%%%%%%%%
There are many possible gauge choices  to study the
perturbations of a stationary, axisymmetric spacetime (see for
instance \cite{SF}). Here we describe a particular gauge which
has the property to reduce, when the background becomes spherically
symmetric, to the well-known Regge-Wheeler gauge \cite{RW}. As
discussed in Section \ref{method}, we assume a background metric in
the coordinates $(t,r,\theta,\phi)$ like (\ref{rotmet2}) or
(\ref{rotmet3}), with $\frac{\partial}{\partial t}$,
$\frac{\partial}{\partial\phi}$ Killing vectors. The metric
perturbations have the form
\begin{equation}
h_{\mu\nu}(t,r,\theta,\phi)=h^{m\,\omega}_{\mu\nu}(r,\theta)
e^{\ii m\phi}e^{-\ii\omega t}\,.
\end{equation}
We shall show  that it is possible to fix the gauge in such a way that
the metric perturbation  takes the form
\begin{equation}
h_{\mu\nu}^{m\,\omega}(r,\theta)=\left(\begin{array}{cc|cc}
e^\nu H^{m\,\omega}_0(r,\theta) & H^{m\,\omega}_1(r,\theta) & 
-\frac{\ii m}{\sin\theta}h^{m\,\omega}_{0}(r,\theta) 
& \sin\theta h^{m\,\omega}_{0,\theta}(r,\theta)\\
\dots & e^\lambda H^{m\,\omega}_2(r,\theta) & 
-\frac{\ii m}{\sin\theta}h^{m\,\omega}_{1}(r,\theta) 
& \sin\theta h^{m\,\omega}_{1,\theta}(r,\theta)\\
\hline \dots & \dots & K^{m\,\omega}(r,\theta)r^2 & 0 \\
\dots & \dots & \dots &  K^{m\,\omega}(r,\theta)r^2\sin^2\theta\\
\end{array}\right)\,,\label{gengauge}
\end{equation}
and depends on  six quantities
\begin{equation}
\left[H_0^{m\,\omega}(r,\theta),\,H_1^{m\,\omega}(r,\theta),\,
H_2^{m\,\omega}(r,\theta),\,K^{m\,\omega}(r,\theta),\,
h_0^{m\,\omega}(r,\theta),\,h_1^{m\,\omega}(r,\theta)\right]\,.
\end{equation}
{\it These quantities  behave as scalars with respect to rotations.}
In order to fix the gauge (\ref{gengauge}) we impose the following conditions
\begin{eqnarray}
&&h_{ab}^{m\,\omega}(r,\theta)\propto\gamma_{ab}\label{propt} ~,\\
&&\int_0^\pi d\theta\sin\theta\left[
Y^{lm}_{,\theta}(\theta,0)h_{A\theta}^{m\,\omega}(r,\theta)+
\frac{\ii m}{\sin^2\theta}Y^{lm}(\theta,0)
h_{A\phi}^{m\,\omega}(r,\theta)\right]=0~~~\forall l\,.\label{intcond}
\end{eqnarray}
These conditions correspond to setting to zero four functions of $(r,\theta)$:
\begin{eqnarray}
&&h_{\theta\phi}^{m\,\omega}(r,\theta)=0\,,\nn\\
&&h_{\theta\theta}^{m\,\omega}(r,\theta)-\frac{1}{\sin^2\theta}
h_{\phi\phi}^{m\,\omega}(r,\theta)=0\,,\nn\\
&&\sum_{l=|m|}^\infty\int_0^\pi d\theta\sin\theta\left[
Y^{lm}_{,\theta}(\theta,0)h_{t\theta}^{m\,\omega}(r,\theta)+
\frac{\ii m}{\sin^2\theta}Y^{lm}(\theta,0)
h_{t\phi}^{m\,\omega}(r,\theta)\right]=0\,,\nn\\
&&\sum_{l=|m|}^\infty\int_0^\pi d\theta\sin\theta\left[
Y^{lm}_{,\theta}(\theta,0)h_{r\theta}^{m\,\omega}(r,\theta)+
\frac{\ii m}{\sin^2\theta}Y^{lm}(\theta,0)
h_{r\phi}^{m\,\omega}(r,\theta)\right]=0\,,
\end{eqnarray}
and can be imposed through a diffeomorphism generated by the vector
field
\begin{equation}
\xi_\mu(t,r,\theta,\phi)=\xi_\mu^{m\,\omega}(r,\theta)
e^{\ii m\phi}e^{-\ii\omega t}\,,
\end{equation}
which depends on four functions of $(r,\theta)$.

The relation between (\ref{gengauge}) and (\ref{propt}) is
trivial, but to show that (\ref{intcond}) implies (\ref{gengauge}) 
is less obvious. If we integrate by parts
(\ref{intcond}), we find
\begin{equation}
\int_0^\pi d\theta Y^{lm}(\theta,0)\left[
-\left(\sin\theta h_{A\theta}^{m\,\omega}(r,\theta)\right)_{,\theta}+
\frac{\ii m}{\sin\theta}
h_{A\phi}^{m\,\omega}(r,\theta)\right]=0\,.
\end{equation}
As it holds for all $l$'s, the term in square brackets must vanish identically, i.e.
\begin{equation}
-\left(\sin\theta h_{A\theta}^{m\,\omega}(r,\theta)\right)_{,\theta}+
\frac{\ii m}{\sin\theta}h_{A\phi}^{m\,\omega}(r,\theta)\equiv0~,
\end{equation}
therefore we can express the four quantities
$\left(h_{A\theta}^{m\,\omega}(r,\theta),
h_{A\phi}^{m\,\omega}(r,\theta)\right)$ in terms of two scalar
functions $h_{A}^{m\,\omega}(r,\theta)$ such that:
\begin{eqnarray}
h_{A\theta}^{m\,\omega}(r,\theta)&=&
-\frac{\ii m}{\sin\theta}h_{A}^{m\,\omega}(r,\theta)\nn\\
h_{A\phi}^{m\,\omega}(r,\theta)&=&\left(
\sin\theta h_{A}^{m\,\omega}(r,\theta)\right)_{,\theta}\,.
\end{eqnarray}
This, together with (\ref{propt}), gives (\ref{gengauge}).  This gauge
choice is implicit in the formulation used in K1 to describe the
perturbations of slowly rotating stars. It can, in principle, also be
chosen to describe perturbations of rapidly rotating stars.

We stress that the existence of the generalized Regge-Wheeler gauge,
in which all perturbations are expressed in terms of functions that
are scalar with respect to rotation, is important because it provides
a solid basis to the approach described in this paper. Indeed, it
guarantees that expanding the perturbations in tensorial spherical
harmonics is equivalent to expand  in circular harmonics $e^{\ii
m\phi}$ firstly, and then to expand the functions appearing in the resulting
equations in $(r,\theta)$, in associate Legendre polynomials,
$P^{lm}(\theta)$.

%%%%%%%%%%%%%%%%%%%%%%%%%%%%%%%%%%%%%%%%%%%%%%%%%%%%%%%%%%%%%%%%%%%%%%%%%%%%
\subsection{Relations with the Regge-Wheeler gauge}
%%%%%%%%%%%%%%%%%%%%%%%%%%%%%%%%%%%%%%%%%%%%%%%%%%%%%%%%%%%%%%%%%%%%%%%%%%%%
In order to better understand how the gauge (\ref{gengauge}) is
related to the Regge-Wheeler (RW) gauge, we now expand
$h_{\mu\nu}(t,r,\theta,\phi)$ in tensor spherical harmonics. This is
always possible (on a surface $t=const.$, $r=const.$), but typically
it is not useful if the background is non-spherical, since the
dynamical equations couple perturbations with different $l$'s. Anyway,
for slowly rotating stars the couplings are small, and the spherical
harmonics expansion, as described in Appendix \ref{couplkoj} and in
\cite{Kojima1992}, turns out to be useful.

By expanding the perturbed metric tensor in tensor spherical harmonics,
before any gauge fixing we find
\begin{eqnarray}
&&h_{\mu\nu}(t,r,\theta,\phi)=h_{\mu\nu}^{m\,\omega}(r,\theta)
e^{\ii m\phi}e^{-\ii\omega t}=\nn\\
&&\sum_{l}\left(\begin{array}{cc|c}
e^\nu H_0^{lm}(r)Y^{lm}(\theta,\phi) & 
H_1^{lm}(r)Y^{lm}(\theta,\phi) & 
h_{0,pol}^{lm}(r)Y_{,a}^{lm}(\theta,\phi)+
h_0^{lm}(r)S_a^{lm}(\theta,\phi) \\
\dots & e^\lambda H_2^{lm}(r)Y^{lm}(\theta,\phi) &  
h_{1,pol}^{lm}(r)Y_{,a}^{lm}(\theta,\phi)+
h_1^{lm}(r)S_a^{lm}(\theta,\phi) \\
\hline
\dots & \dots & K^{lm}(r)r^2\gamma_{ab}Y^{lm}(\theta,\phi) + 
G^{lm}(r)Z_{ab}^{lm}(\theta,\phi)
+h_{ax}^{lm}(r)S_{ab}^{lm}(\theta,\phi)\\
\end{array}\right)e^{-\ii\omega t}\nn\\
\end{eqnarray}
where
\begin{equation}
S_a^{lm}(\theta,\phi)=(S^{lm}_\theta,S^{lm}_\phi)=
\left(-\frac{1}{\sin\theta}Y^{lm}_{,\phi}
,\sin\theta Y^{lm}_{,\theta}\right)
\end{equation}
are the axial vector harmonics, and $Z_{ab}$ and $S_{ab}$ are tensor
harmonics satisfying $\gamma^{ab}Z_{ab}=\gamma^{ab}S_{ab}=0$,
with polar and axial parity, respectively.

The RW-gauge for a spherical background, imposes
\cite{RW}:
\begin{equation}
h_{0,pol}^{lm}=h_{1,pol}^{lm}=h_{ax}^{lm}=G^{lm}=0\,.\label{RWgauge}
\end{equation}
If we consider (\ref{RWgauge}) in the case of a non-spherical
background, we see that it is equivalent to the gauge
(\ref{gengauge}). Indeed, expanding
$h_{Aa}(t,r,\theta,\phi)$ in vector spherical
harmonics, we find
\begin{equation}
h_{Aa}^{m\,\omega}(r,\theta)e^{\ii m\phi}=\sum_{l}\left[
h_{A\,pol}^{lm}(r)Y_{,a}^{lm}(\theta,\phi)+
h_{A\,ax}^{lm}(r)S_a^{lm}(\theta,\phi)\right]\,,
\end{equation}
which, if $h_{A\,pol}^{lm}=0$, reduces to
\begin{eqnarray}
h_{Aa}^{m\,\omega}(r,\theta)e^{\ii m\phi}&=&\sum_{l}
h_{A\,ax}^{lm}(r)S_a^{lm}(\theta,\phi)=\sum_lh_{A\,ax}^{lm}(r)
\left(-\frac{1}{\sin\theta}Y^{lm}_{,\phi}
,\sin\theta Y^{lm}_{,\theta}\right)\nn\\
&=&\left(-\frac{\ii m}{\sin\theta}h_A,\sin\theta h_{A,\theta}\right)
e^{\ii m\phi}\,,
\end{eqnarray}
where we have defined the scalar functions
\begin{equation}
h_A(r,\theta)\equiv\sum_{l}h_{A\,ax}^{lm}(r)Y^{lm}(\theta,0)\,.
\label{defhA}
\end{equation}
%%%%%%%%%%%%%%%%%%%%%%%%%%%%%%%%%%%%%%%%%%%%%%%%%%%%%%%%%%%%%%%%%%%%%%%%%%%%
\subsection{The Regge-Wheeler equation for a spherical star as a 2D-equation in $r$ and
$\theta$}
\label{RWspher2d}
%%%%%%%%%%%%%%%%%%%%%%%%%%%%%%%%%%%%%%%%%%%%%%%%%%%%%%%%%%%%%%%%%%%%%%%%%%%%
We conclude this section with a simple exercise.  Choosing the gauge
(\ref{gengauge}), we derive the equations which describe the axial
perturbations of a non rotating star, i.e. the equations for
$h^{m\,\omega}_A(r,\theta)$ (in this case fluid perturbations are
decoupled from metric perturbations), and we show how to solve them
using spectral methods.  We shall follow the lines of the well-known
derivation of the Regge-Wheeler equation \cite{RW}, with one
difference: we do not expand the perturbations in spherical harmonics.
By defining the function
\begin{equation}
Z^{m\,\omega}(r,\theta)\equiv\frac{1}{r}
h^{m\,\omega}_1(r,\theta)e^{(\nu-\lambda)/2}\,,
\label{deffZ}
\end{equation}
from Einstein's equations we find
\begin{equation}
h^{m\,\omega}_0(r,\theta)=\frac{1}{\ii\sigma}(rZ^{m\,\omega}(r,\theta))'
e^{(\nu-\lambda)/2}\,,
\end{equation}
where `'' indicates differentiation with respect to $r$.
$Z^{m\,\omega}(r,\theta)$ satisfies the partial differential equation
\begin{equation}
\partial^2_{r_*}Z^{m\,\omega}
+(\sigma^2-V)Z^{m\,\omega}=0
\label{RW2d}
\end{equation}
where the coordinate $r_*$ is defined by 
\begin{equation}
\frac{dr_*}{dr}=e^{(\lambda-\nu)/2}\,.\label{tortoise}
\end{equation}
In this formulation,  $V$ is a differential operator:
\begin{equation}
V\equiv\frac{e^{\nu}}{r^2}\left[
-\left(\partial^2_{\theta}+\cot\theta\partial_\theta-
\frac{m^2}{\sin^2\theta}\right)-\frac{6M}{r}+
4\pi(\rho-p)r^2\right]\,.\label{V2d}
\end{equation}
The usual one-dimensional Regge-Wheeler equation can easily be
recovered by expanding $Z^{m\,\omega}(r,\theta)$ in scalar spherical
harmonics:
\begin{equation}
Z^{m\,\omega}(r,\theta)=\sum_{l=l_{min}}^\infty 
Z^{lm\,\omega}(r)Y^{lm}(\theta,0)\label{ZY}
\quad\rightarrow\quad
\begin{cases}
\frac{d^2}{dr_*^2}Z^{lm\,\omega}(r)
+(\sigma^2-V^l)Z^{lm\,\omega}(r)=0&\cr
V^l=\frac{e^{\nu}}{r^2}\left(l(l+1)-
\frac{6M}{r}+4\pi(\rho-p)r^2\right)&\cr
\end{cases}
\end{equation}
where
\begin{equation}
l_{min}\equiv{\rm max}(|m|,2)\,,
\end{equation}
and $Z^{lm\,\omega}(r)$ is the standard Regge-Wheeler function.

Let us briefly describe how equation (\ref{RW2d}) can be integrated
using spectral methods and the standing wave approach. We shall
integrate this equation for real values of the frequency, therefore in
the following we shall set $\omega=\sigma$.  The integration range in
$r_*$ is $r_*=[r_{*1},r_{*2}]$, i.e. given
$Z^{m\,\sigma}(r_{*1},\theta)$ we want to know
$Z^{m\,\sigma}(r_{*2},\theta)$. The starting point $r_{*1}$
corresponds to a small sphere near the center of the star, with
$r=r_0\ll R$, where the Regge-Wheeler function is given by an
analytical expansion (see below). The final point, $r_{*2}$,
corresponds to a point in the wave zone, $r=r_\infty\gg R$, where the
ingoing and outgoing amplitudes can be extracted.

We rescale the variables $r_*$ and $\theta$ as follows:
\begin{eqnarray}
x&=&\frac{2r_*-r_{*1}-r_{*2}}{r_{*1}-r_{*2}}\in[-1,1]\,\nn\\
y&=&\cos\theta\in[-1,1]\,,
\end{eqnarray}
so that $r_{*1}$ corresponds to $x=1$, and $r_{*2}$ to $x=-1$. Then,
we perform the double expansion (\ref{doubleexp}) of the Regge-Wheeler
function $Z^{m\,\sigma}(r,\theta)$ defined in (\ref{deffZ}), for
assigned values of $m$, $\sigma$:
\begin{equation}
Z^{m\,\sigma}(x,y)=\sum_{n=0}^K\sum_{l=l_{min}}^La^{lm\,\sigma}_n
T_n(x)P^{lm}(y)\,.
\end{equation}
The expansions in Chebyshev's and associated Legendre's polynomials
are truncated at $K$ and $L$, respectively (for instance, $L=10,K=20$).  The
boundary conditions at the center of the star are imposed by assigning
$Z^{m\,\sigma}$ and its derivative at $x=1$:
\begin{eqnarray}
Z^{m\,\sigma}(x=1,y)&=&\sum_{l=l_{min}}^Lz^l_0\hat H^{lm}P^{lm}(y)\nn\\
\pa_{r_*}Z^{m\,\sigma}(x=1,y)&=&\sum_{l=l_{min}}^Lz^l_1\hat H^{lm}P^{lm}(y)\,,
\end{eqnarray}
where the analytic expansion of the Regge-Wheeler equation gives
\cite{chandrafer}
\begin{eqnarray}
z^l_0&=& r_0^{l+1}+X^l_{in}(\sigma)r_0^{l+3}\nn\\
z^l_1&=& e^{\nu(0)}
\left[(l+1)r_0^l+(l+3)X^l_{in}(\sigma)r_0^{l+2}\right]\nn\\
X^l_{in}(\sigma)&\equiv&\frac{(l+2)\left[\frac{1}{3}(2l-1)
\rho(0)-p(0)\right]-\sigma^2e^{\nu(0)}}{2(2l+3)}
\end{eqnarray}
(note that while in $z_{0,1}^l$  and $X^l_{in}$, the index $l$ is a
superscript, in $r_0^l$ it is an exponent).

The constants $\hat H^{lm}$ form a vector 
\begin{equation}
{\bf \hat H}^m\equiv\left(\hat H^{lm}\right)\,,
\end{equation}
which can be freely assigned; for each vector ${\bf\hat H}$ we have
one solution of the equation. Notice that we have one constant $\hat
H^{lm}$ for each value of $l$ i.e., in the language of Section
\ref{standingrot}, $N=1$ for the axial parity perturbations.  If, in
addition to axial perturbations, polar parity perturbations are
considered, described outside the star by the Zerilli equation, then
there is another constant to be assigned for each value of
$l$. Therefore, if all metric perturbations are considered, $N=2$ as
discussed in Section \ref{standingrot}.

We now project equation (\ref{RW2d}) in the basis of Chebyshev's and
associate Legendre's polynomials. The operator
\begin{equation}
\partial^2_{\theta}+\cot\theta\partial_\theta-\frac{m^2}{\sin^2\theta}
\end{equation}
is diagonal in the $P^{lm}$ basis, with eigenvalue
$-l(l+1)$. Therefore, in this basis the operator $V$ defined in
(\ref{V2d}) reduces to the one-dimensional Regge-Wheeler potential
$V^l(r)$ (\ref{ZY}).  We introduce (as in in Section \ref{cheby},
Equations (\ref{defder}), (\ref{Vnm})) the derivative matrix $D_{nn'}$
and the potential matrix $V_{nm}$ obtained projecting $V^l(r)$ on
Chebyshev polynomials.  To write the matrix equation we define the
collective index
\begin{equation}
i(l,n)=(l-|m|)(K+1)+n+1\in[1,{\cal N}]
\end{equation}
with ${\cal N}=(L-|m|+1)(K+1)$, and we define the ${\cal N}$-dimensional
vector of components
\begin{equation}
\hat a^m_i=\hat a^m_{i(l,n)}=a^{lm}_n\,.
\end{equation}
The matrix equation has the block-diagonal form
\begin{equation}
\hat L_{ii'}a^m_{i'}=0\label{La2d}
\end{equation}
with
\begin{equation}
\hat L=\left(\begin{array}{cccc}
\left(D^2_{nn'}+\tilde V_{nn'}\right)_{l=|m|} &  & & 0 \\  & 
\left(D^2_{nn'}+\tilde V_{nn'}\right)_{l=|m|+1} &  &  \\
 & & \ddots & \\ 0 &  &  & \left(D^2_{ll'}+\tilde V_{nn'}\right)_{l=L}\\
\end{array}\right)\,.\label{matrixL}
\end{equation}
The boundary conditions are implemented by replacing the last two
lines of each block, i.e. each $i(l,K)$-th and $i(l,K-1)$-th lines,
with the conditions at the center in terms of the vector ${\bf\hat
H}^m=(\hat H^{lm})$:
\begin{eqnarray}
\sum_{n=0}^K\hat a^m_{i(l,n)}&=&z^l_0\hat H^{lm}\nn\\
\sum_{n,k}^{0,K}\hat D_{kn}\hat a^m_{i(l,n)}&=&z^l_1\hat H^{lm}\,.\label{boundRW2d}
\end{eqnarray}
By inverting the matrix equation (\ref{La2d}) we find the coefficients
$\hat a^m_i$ and then $Z^{m\,\omega}(-1,y)$.

In this way, for each choice of $(m,\sigma)$, and of the vector ${\bf
\hat H}^m$ of initial conditions, we can integrate from $r_0$ to
$r_\infty$.  The ingoing amplitude at infinity
\begin{equation}
{\cal A}^m(\sigma,\theta)=\sum_l{\cal A}^{lm}(\sigma)P^{lm}(\theta)\,.
\end{equation}
can be found using the algorithm described in \cite{SW2}.  Therefore,
for each choice of the vector of initial conditions ${\bf\hat
H}^m=(\hat H^{lm})$, we have the vector of ingoing amplitudes at
infinity
\begin{equation}
{\bf A}^m=({\cal A}^{lm}(\sigma))\,,
\end{equation}
This procedure  is linear at any step, thus, repeating
it for all ${\bf\hat H}^m$:
\begin{equation}
{\bf\hat H}^m=(1,0,\dots,0)\,,~~~~~{\bf\hat H}^m=(0,1,\dots,0)\,,~~~~~\dots,
~~~~~{\bf\hat H}^m=(0,0,\dots,1)
\end{equation}
we find the expression of the matrix ${\bf M}^m=({\cal
M}^{m|ll'}(\sigma))$ defined by
\begin{equation}
{\cal A}^{lm}(\sigma)={\cal M}^{m|ll'}(\sigma)\hat H^{l'm}\,.
\end{equation}
As explained in Section \ref{standingrot}, near a mode
$\sigma\sim\sigma_0$ the modulus of the determinant of this matrix
behaves as
\begin{equation}
|\det{\bf M}^m(\sigma)|\propto
\sqrt{(\sigma-\sigma_0)^2+\frac{1}{\tau_0}^2}
\end{equation}
and the frequency $\sigma_0$ and the damping time $\tau_0$ of the mode
can be found by a quadratic fit in $\sigma$.

%%%%%%%%%%%%%%%%%%%%%%%%%%%%%%%%%%%%%%%%%%%%%%%%%%%%%%%%%%%%%%%%%%%%%%%%%%%%
\section{Equations for the perturbations of slowly rotating stars}
\label{couplkoj}
%%%%%%%%%%%%%%%%%%%%%%%%%%%%%%%%%%%%%%%%%%%%%%%%%%%%%%%%%%%%%%%%%%%%%%%%%%%%
This derivation of the equations describing the perturbations of a
slowly rotating star is based on the work of Kojima \cite{Kojima1992},
denoted as K1. As discussed above, since we are considering slowly
rotating stars, we can first expand the perturbations in spherical
harmonics, getting a system of coupled ODE in $r$, and then expand
this system in Chebyshev polynomials, getting an algebraic matrix
equation. The first step of this program is equivalent to the
derivation of K1, with one difference: we are going to reformulate the
equations in terms of a different set of variables, which are
numerically well behaved near the center of the star. Following K1, we
shall assume $l\ge2$.

The background configuration, describing a slowly rotating, stationary
and axially symmetric star is given by equation (\ref{slowstar}).  The
pressure $p$ and the energy density $\rho$ are found by solving the
TOV equations; we assume that the equation of state of matter in the
star is barotropic, $p=p(\rho)$; therefore,
$c_s^2\equiv\left(\frac{\partial
p}{\partial\rho}\right)=\frac{p'}{\rho'}$.
The perturbations of the background (\ref{slowstar}) can be written as
\begin{eqnarray}
&&h_{\mu\nu}=\sum_{lm}\left(\begin{array}{cc|cc}
e^\nu H^{lm}_0(r)Y^{lm}(\theta,\!\phi) & 
\!\!H^{lm}_1(r)Y^{lm}(\theta,\!\phi) & 
h^{lm}_{0}(r)S_\theta^{lm}(\theta,\!\phi) 
& h^{lm}_{0}(r)S_\phi^{lm}(\theta,\!\phi)\\
\dots & \!\!\!\!\!\!\!\!e^\lambda H^{lm}_2(r)
Y^{lm}(\theta,\!\phi) & h^{lm}_{1}(r) 
S_\theta^{lm}(\theta,\!\phi)& h^{lm}_1(r)S_\phi^{lm}(\theta,\!\phi)\\
\hline \dots & \dots & K^{lm}(r)r^2Y^{lm}(\theta,\!\phi) & 0 \\
\dots & \dots & \dots &  \!\!\!\!\!\!K^{lm}(r)r^2\sin^2\theta 
Y^{lm}(\theta,\!\phi)\\
\end{array}\right)e^{-\ii\sigma t}\nn\\
&&\delta u^r=\frac{e^{\nu/2-\lambda}}{4\pi(\rho+p)}
R^{lm}(r)Y^{lm}(\theta,\!\phi)e^{-\ii\sigma t}\nn\\
&&\delta u^\theta=\frac{e^{\nu/2-\lambda}}{4\pi(\rho+p)}\left(
V^{lm}(r)Y^{lm}_{,\theta}(\theta,\!\phi)
+U^{lm}(r)S^{lm}_\theta(\theta,\!\phi)\right)e^{-\ii\sigma t}\nn\\
&&\delta u^\phi=\frac{e^{\nu/2-\lambda}}{4\pi(\rho+p)}\left(
V^{lm}(r)Y^{lm}_{,\phi}(\theta,\!\phi)+U^{lm}(r)S^{lm}_\phi(\theta,\!\phi)
\right)e^{-\ii\sigma t}\nn\\
&&\delta\rho=\delta\rho^{lm}(r)Y^{lm}(\theta,\!\phi)e^{-\ii\sigma t}\nn\\
&&\delta p=\delta p^{lm}(r)Y^{lm}(\theta,\!\phi)e^{-\ii\sigma t}\,,
\label{pertmetric2}
\end{eqnarray}
with $\sigma$ real. The linearized Einstein equations for the radial
part of these quantities are ordinary differential equations in $r$.
As explained in Section \ref{slowly}, the general structure of these
equations is:
\begin{eqnarray}
{\cal L}^{pol}[H_0^{lm},K^{lm};\sigma]
&=&m{\cal E}[H_0^{lm},K^{lm};\sigma]
+{\cal F}^{(\pm)}[Z_{RW}^{l\pm1\,m};\sigma]\nn\\
{\cal L}^{ax}[Z_{RW}^{lm};\sigma]&=&
m{\cal N}[Z_{RW}^{lm};\sigma]+
{\cal D}^{(\pm)}[H_0^{l\pm1\,m},K^{l\pm1\,m};\sigma]\,.\label{kojstruct2}
\end{eqnarray}
The quantities $H_1^{lm}$, $H_2^{lm}$, $R^{lm}$, $V^{lm}$, $U^{lm}$,
$\delta\rho^{lm}$, $\delta p^{lm}$ can be expressed in terms of
$H_0^{lm},\,K^{lm},\,Z_{RW}^{lm}$, once equations (\ref{kojstruct2})
have been solved.
%%%%%%%%%%%%%%%%%%%%%%%%%%%%%%%%%%%%%%%%%%%%%%%%%%%%%%%%%%%%%%%%%
\subsection{The $O(\epsilon^0)$ equations}
%%%%%%%%%%%%%%%%%%%%%%%%%%%%%%%%%%%%%%%%%%%%%%%%%%%%%%%%%%%%%%%%%
The equations at $O(\epsilon^0)$ describe the perturbations of a
non-rotating star.

The equations for polar perturbations inside the star are a system of
two second order ODE in $K^{lm}$, $H_0^{lm}$:
\begin{eqnarray}
L_{int}^{(1)lm}[H_0,K]&\equiv&(K^{lm}-H_0^{lm})''
-\frac{e^\lambda}{r^2}[2r-10M+4\pi(\rho-5p)r^3](K^{lm}-H_0^{lm})'\nn\\
&&+\frac{e^\lambda}{r^2}(\sigma^2 e^{-\nu}r^2-2n)(K^{lm}-H_0^{lm})\nn\\
&&+4\frac{e^\lambda}{r^4}[3Mr-4\pi\rho r^4-e^\lambda(M+4\pi pr^3)^2]
H_0^{lm}=0\label{in1norot}\\
L_{int}^{(2)lm}[H_0,K]&\equiv&K^{lm\,''}
-\frac{e^\lambda}{r^2}[(r-3M-4\pi pr^3)c_s^{-2}-3r+5M+4\pi\rho r^3]
K^{lm\,'}\nn\\
&&+\frac{e^\lambda}{r^2}[\sigma^2e^{-\nu}r^2c_s^{-2}-n(c_s^{-2}+1)]K^{lm}
+\frac{c_s^{-2}-1}{r}H_0^{lm\,'}\nn\\
&&+\frac{e^\lambda}{r^3}[(nr+4M+8\pi pr^3)c_s^{-2}-(n+2)r
+8\pi\rho r^3]H_0^{lm}=0\,.\label{in2norot}
\end{eqnarray}

Once $H_0^{lm}$, $K^{lm}$ have been determined, $H_1^{lm}$ and
$H^{lm}_2$ can be computed through the following relations:
\begin{eqnarray}
H^{lm}_1&=&-\frac{e^\nu}{\ii\sigma}\left(H_0^{lm\,'}-K^{lm\,'}
+\frac{2e^\lambda}{r^2}(M+4\pi pr^3)H_0^{lm}\right)\nn\\
H^{lm}_2&=&H^{lm}_0\,.\label{H12rel}
\end{eqnarray}
Other relations (see K1) give the fluid perturbations $R^{lm}$,
$V^{lm}$, $U^{lm}$, $\delta\rho^{lm}$, $\delta p^{lm}$ in terms of the
metric perturbations.

Outside the star, the polar perturbations reduce to a system of
three first order ODE in $K^{lm}$, $H_1^{lm}$ and $H_0^{lm}$:
\begin{eqnarray}
L_{ext}^{(1)lm}[H_0,\tilde H_1,K]&\equiv&
K^{lm\,'}+\frac{e^\lambda(r-3M)}{r^2}K^{lm}-\frac{1}{r}H_0^{lm}
-\frac{n+1}{r}\tilde H_1^{lm}=0\label{out1norot}\\
L_{ext}^{(2)lm}[H_0,\tilde H_1,K]&\equiv&
-r\tilde H_1^{lm\,'}+e^\lambda(K^{lm}+H_0^{lm})-
\frac{2Me^\lambda}{r}\tilde H_1^{lm}=0\label{out2norot}\\
L_{ext}^{(3)lm}[H_0,\tilde H_1,K]&\equiv&
H_0^{lm\,'}+\frac{e^\lambda(r-3M)}{r^2}K^{lm}
-\frac{e^\lambda(r-4M)}{r^2}H_0^{lm}\nn\\
&&+\left(\sigma^2 re^\lambda-\frac{n+1}{r}\right)\tilde H_1^{lm}=0\,,
\label{out3norot}
\end{eqnarray}
where we have defined
\begin{equation}
\tilde H_1^{lm}\equiv-\frac{H_1^{lm}}{\ii\sigma r}
\end{equation}
in order to have equations with real coefficients.
Furthermore, there is an algebraic constraint:
\begin{eqnarray}
L_{ext}^{(4)lm}[H_0,\tilde H_1,K]&\equiv&
(\sigma^2r^4e^\lambda-nr^2-Mr+M^2e^\lambda)K^{lm}\nn\\
&&+(nr+3M)rH_0^{lm}-[\sigma^2r^4-(n+1)Mr]\tilde H_1^{lm}=0\,.\nn\\
\label{algebraicnorot}
\end{eqnarray}
Equations (\ref{out1norot})-(\ref{out3norot}) are equivalent to the
Zerilli equation, but they have the advantage to be easily
generalizable to rotating stars, as we will see below. The Zerilli
function $Z_{Zer}$ can be computed in terms of $K$, $\tilde H_1$:
\begin{equation}
Z_{Zer}^{lm}=\frac{r^2}{nr+3M}(K^{lm}-e^\nu\tilde H_1^{lm})\,.\label{defzer}
\end{equation}
The axial perturbations are described by the Regge-Wheeler equation
\begin{equation}
L^{lm}[Z_{RW}]\equiv 
\frac{d^2}{dr_*^2}Z_{RW}^{lm}
+\left[\sigma^2-e^\nu
\left(\frac{l(l+1)}{r^2}-\frac{6M}{r^3}+4\pi(\rho-p)\right)\right]
Z_{RW}^{lm}=0\,,\label{RWinnorot}
\end{equation}
where the coordinate $r_*$ has been defined in (\ref{tortoise}) and
\begin{eqnarray}
h_0^{lm}&=&-\frac{e^{(\nu-\lambda)/2}}{\ii\sigma}\left(Z^{lm}_{RW}r
\right)'\nn\\
h_1^{lm}&=&e^{(\lambda-\nu)/2}Z^{lm}_{RW}r\,.
\end{eqnarray}
An analytical expansion of equations (\ref{in1norot}),
(\ref{in2norot}), (\ref{RWinnorot}) near the center of the star gives, 
for each value of $l$, three independent 
conditions at the center:
\begin{eqnarray}
K^{lm}-H_0^{lm}&=&r^{l+2}+kh^{lm(+)}r^{l+4}\nn\\
K^{lm}&=&r^l+k^{lm(+)}r^{l+2}\nn\\
Z_{RW}^{lm}&=&0\label{condcenter1}\\
&&\nn\\
K^{lm}-H_0^{lm}&=&r^{l+2}+kh^{lm(-)}r^{l+4}\nn\\
K^{lm}&=&-r^l+k^{lm(-)}r^{l+2}\nn\\
Z_{RW}^{lm}&=&0\label{condcenter2}\\
&&\nn\\
K^{lm}-H_0^{lm}&=&0\nn\\
K^{lm}&=&0\nn\\
Z_{RW}^{lm}&=&r^{l+1}+z^{lm}r^{l+3}\label{condcenter3}
\end{eqnarray}
where the expressions for $kh^{lm(\pm)}$, $k^{lm(\pm)}$, $z^{lm}$ can
be evaluated from the analytical expansion.

We notice that $K^{lm}$ and $H_0^{lm}$ behave, as $r\rightarrow0$,
like $r^l$, while the combination $K^{lm}-H_0^{lm}$ behaves as
$r^{l+2}$. Consequently, when we expand $K^{lm}$ and $H_0^{lm}$ in
powers of $r$ about $r=0$, we find that the leading terms are
coincident. In other words, the differential equations for the
variables $K^{lm}$, $H_0^{lm}$ and $Z_{RW}^{lm}$ are linearly
dependent near the origin (see the discussion in \cite{chandrafer}).
To avoid this problem, we use as integration variables
$K^{lm}-H_0^{lm}$, $K^{lm}$ and $Z_{RW}^{lm}$.

Finally, we notice that at order $O(\epsilon^0)$, i.e. for a non
rotating star, equation (\ref{H12rel}) establishes that
$H^{lm}_2=H^{lm}_0$.  Therefore it is equivalent to use either
$H_0^{lm}$ or $H_2^{lm}$.

%%%%%%%%%%%%%%%%%%%%%%%%%%%%%%%%%%%%%%%%%%%%%%%%%%%%%%%%%%%%%%%%%
\subsection{The $O(\epsilon^1)$ equations}
%%%%%%%%%%%%%%%%%%%%%%%%%%%%%%%%%%%%%%%%%%%%%%%%%%%%%%%%%%%%%%%%%
We define
\begin{equation}
\begin{array}{ccc}
n_+\equiv\frac{l(l+3)}{2}&~~~~~~~~&n_-\equiv\frac{(l-2)(l+1)}{2}\\
&&\\
Q_{l-1\,m}\equiv\sqrt{\frac{(l-m)(l+m)}{(2l-1)(2l+1)}}&&
Q_{l+1\,m}\equiv\sqrt{\frac{(l+1-m)(l+1+m)}{(2l+1)(2l+3)}}\,.\\
\end{array}
\end{equation}
The perturbed equations  inside the star have the form
\begin{eqnarray}
L_{int}^{(J)lm}[H_0,K]&=&-\frac{m}{(n+1)\sigma}
E^{(J)lm}[H_0,K]\nn\\
&&+e^{-(\lambda+\nu)/2}
\left[\frac{\ii Q_{l-1\,m}D^{(J)l-1\,m}[Z_{RW}]}{\sigma(n-n_-)}+
\frac{\ii Q_{l+1\,m}D^{(J)l+1\,m}[Z_{RW}]}{\sigma(n-n_+)}\right]\nn\\
L^{lm}[Z_{RW}]&=&\frac{m}{\sigma}N^{lm}[Z_{RW}]+e^{(\lambda+3\nu)/2}
\left[\frac{\ii Q_{l-1\,m}F^{l-1\,m}[H_0,K]}{\sigma(n-n_-)}+
\frac{\ii Q_{l+1\,m}F^{l+1\,m}[H_0,K]}{\sigma(n-n_+)}\right]
\label{inrot0}
\end{eqnarray}
($J=1,2$), where $L_{int}^{(J)lm}[H_0,K]$, $L^{lm}[Z_{RW}]$ are the
operators defined in equations (\ref{in1norot}), (\ref{in2norot}),
(\ref{RWinnorot}) and $E^{(J)lm}[H_0,K]$, $D^{(J)l\pm1\,m}[Z_{RW}]$,
$N^{lm}[Z_{RW}]$, $F^{l\pm1\,m}[H_0,K]$ are operators at first order
in $\epsilon$, whose explicit expressions are given in K1. The
operators $D^{(J)l\pm1\,m}[Z_{RW}]$, $F^{l\pm1\,m}[H_0,K]$ couple
perturbations belonging to different $l$'s, and were neglected in the
numerical integration of K2.

In order to have equations with real coefficients, we need to get rid
of the factors $\ii$ in (\ref{inrot0}). To this purpose, we redefine
the Regge-Wheeler function by a factor $-\ii$
\begin{equation}
Z^{lm}_{RW}\Rightarrow-\ii Z^{lm}_{RW}\,,
\end{equation}
thus equations (\ref{inrot0}) become
\begin{eqnarray}
L_{int}^{(J)lm}[H_0,K]&=&-\frac{m}{(n+1)\sigma}
E^{(J)lm}[H_0,K]\nn\\
&&+e^{-(\lambda+\nu)/2}
\left[\frac{Q_{l-1\,m}D^{(J)l-1\,m}[Z_{RW}]}{\sigma(n-n_-)}+
\frac{Q_{l+1\,m}D^{(J)l+1\,m}[Z_{RW}]}{\sigma(n-n_+)}\right]\nn\\
L^{lm}[Z_{RW}]&=&\frac{m}{\sigma}N^{lm}[Z_{RW}]-e^{(\lambda+3\nu)/2}
\left[\frac{Q_{l-1\,m}F^{l-1\,m}[H_0,K]}{\sigma(n-n_-)}+
\frac{Q_{l+1\,m}F^{l+1\,m}[H_0,K]}{\sigma(n-n_+)}\right]\,.
\label{inrot}
\end{eqnarray}
This rescaling consistently eliminates all imaginary units from the
equations.

The numerical integration of equations (\ref{inrot}) presents a
serious problem.  If we perform an analytical expansion near the
center of (\ref{inrot}), we find that the coupling terms
$D^{(J)l-1\,m}[Z_{RW}^{lm}]$ become larger than
$L_{int}^{(J)lm}[H_0^{lm},K^{lm}]$ as $r\rightarrow0$. The reason
behind this pathological behaviour is that near the center of the star
\begin{eqnarray}
K^{lm}-H_0^{lm}&\sim&r^{l+2} + (~O(\epsilon)~{\rm terms}~)\cdot r^l\nn\\
K^{lm}&\sim&r^l\nn\\
H_0^{lm}&\sim&r^l\,.
\end{eqnarray}
If we use as integration variable $H_2^{lm}$ instead of $H_0^{lm}$
(notice that while in the non rotating case $H_0^{lm}=H_2^{lm}$, if
the star rotates $H_0^{lm}=H_2^{lm}+O(\epsilon)$) this problem is
overcome, since
\begin{equation}
K^{lm}-H_2^{lm}\sim r^{l+2}\,.
\end{equation}
Consequently, the coupling terms in the perturbed equations are
smaller than $L_{int}^{(J)lm}[H_2^{lm},K^{lm}]$. For this reason, we
have expressed our equations inside the star in terms of
$K^{lm}-H_2^{lm}$, $K^{lm}$, $Z_{RW}^{lm}$:
\begin{eqnarray}
L_{int}^{(J)lm}[H_2,K]&=&-\frac{m}{(n+1)\sigma}
\tilde E^{(J)lm}[H_2,K]\nn\\
&&+e^{-(\lambda+\nu)/2}
\left[\frac{Q_{l-1\,m}\tilde D^{(J)l-1\,m}[Z_{RW}]}{\sigma(n-n_-)}+
\frac{Q_{l+1\,m}\tilde D^{(J)l+1\,m}[Z_{RW}]}{\sigma(n-n_+)}\right]\nn\\
L^{lm}[Z_{RW}]&=&\frac{m}{\sigma}N^{lm}[Z_{RW}]-e^{(\lambda+3\nu)/2}
\left[\frac{Q_{l-1\,m}F^{l-1\,m}[H_2,K]}{\sigma(n-n_-)}+
\frac{Q_{l+1\,m}F^{l+1\,m}[H_2,K]}{\sigma(n-n_+)}\right]\,.
\label{inrotnew}
\end{eqnarray}
The operators $\tilde E^{(J)lm}$, $\tilde D^{(J)l\pm1\,m}$ are
different from $E^{(J)lm}$, $D^{(J)l\pm1\,m}$ given in K1. Their
expressions are the following:
\begin{eqnarray}
\tilde E^{(1)lm}[H_2,K]&=&\frac{\ii\sigma}{2}\left[2f^{lm\,\prime\prime}(r)
+\left(4\nu'-\lambda'-\frac{6}{r}\right)f^{lm\,\prime}(r)\right.\nn\\
&&\left.-\left(\frac{2e^\lambda}{r^2}(n+1)-(\nu')^2+\frac{4}{r}\left(2\nu'-
\lambda'-\frac{2}{r}\right)-32\pi pe^\lambda\right)f^{lm}(r)\right]\nn\\
&&+\ii\sigma\left(2n\xi^{(1)lm\,\prime}(r)-\tilde\beta^{(1)lm\,\prime}(r)
-\zeta^{(1)lm\,\prime}(r)\right)\nn\\
&&-\frac{\ii\sigma}{2}\left(\frac{2}{r}+\lambda'-2\nu'\right)
\left(2n\xi^{(1)lm}(r)-\tilde\beta^{(1)lm}(r)-\zeta^{(1)lm}(r)\right)\nn\\
&&+\ii\sigma(n+1)e^\lambda C^{(3)lm}(r)-\ii\sigma(n+1)C^{(2)lm}(r)
\label{E1lm}\\
\tilde E^{(2)lm}[H_2,K]&=&-\frac{\ii\sigma}{r}c_s^{-2}\left[f^{lm\,\prime}(r)
+\left(\nu'-\frac{2}{r}+(n+1)\frac{e^\lambda}{r}\right)f^{lm}(r)\right]\nn\\
&&-\frac{\ii\sigma}{r}c_s^{-2}\left(2n\xi^{(1)lm}(r)-\tilde\beta^{(1)lm}(r)
-\zeta^{(1)lm}(r)\right)\nn\\
&&-\frac{\ii\sigma}{2}(n+1)c_s^{-2}C^{(2)lm}(r)+
\frac{\ii\sigma}{2}(n+1)e^{\lambda-\nu}C^{(0)lm}(r)\label{E2lm}\\
\tilde D^{(1)l\pm1\,m}[Z_{RW}]&=&\ii\sigma
e^{(\lambda+\nu)/2}\left[2g^{l\pm1\,m\,\prime\prime}(r)
+\left(4\nu'-\lambda'-\frac{6}{r}\right)g^{l\pm1\,m\,\prime}(r)\right.\nn\\
&&\left.-\left(\frac{2e^\lambda}{r^2}(n+1)-(\nu')^2+\frac{4}{r}\left(2\nu'-
\lambda'-\frac{2}{r}\right)-32\pi pe^\lambda\right)g^{l\pm1\,m}(r)\right]\nn\\
&&+2\ii\sigma e^{(\lambda+\nu)/2}\left[-2(n-2n_\pm-2)
\chi^{(1)l\pm1\,m\,\prime}(r)+(n-n_\pm-1)\tilde\alpha^{(1)l\pm1\,m\,\prime}(r)
-\eta^{(1)l\pm1\,m\,\prime}(r)\right]\nn\\
&&-\ii\sigma e^{(\lambda+\nu)/2}\left(\frac{2}{r}+\lambda'
-2\nu'\right)\left[-2(n-2n_\pm-2)\chi^{(1)l\pm1\,m}(r)\right.\nn\\
&&\left.+(n-n_\pm-1)\tilde\alpha^{(1)l\pm1\,m}(r)-\eta^{(1)l\pm1\,m}(r)
\right]\nn\\
&&+\ii\sigma e^{(3\lambda+\nu)/2}\left[(n-n_\pm)\tilde
A^{(3)l\pm1\,m}(r)+(n-n_\pm)(n-n_\pm-1)B^{(3)l\pm1\,m}(r)
\right]\nn\\
&&-\ii\sigma e^{(\lambda+\nu)/2}\left[(n-n_\pm)\tilde
A^{(2)l\pm1\,m}(r)+(n-n_\pm)(n-n_\pm-1)B^{(2)l\pm1\,m}(r)
\right]\label{D1lm}\\
\tilde D^{(2)l\pm1\,m}[Z_{RW}]&=&-\ii\sigma
\frac{2e^{(\lambda+\nu)/2}}{r}c_s^{-2}
\left[g^{lm\,\prime}(r)+\left(\nu'-\frac{2}{r}+(n+1)
\frac{e^\lambda}{r}\right)g^{lm}(r)\right]\nn\\
&&-\ii\sigma\frac{2e^{(\lambda+\nu)/2}}{r}c_s^{-2}
\left[-2(n-2n_\pm-2)\chi^{(1)l\pm1\,m}(r)\right.\nn\\
&&\left.+(n-n_\pm-1)\tilde\alpha^{(1)l\pm1\,m}(r)-\eta^{(1)l\pm1\,m}(r)
\right]\nn\\
&&-\ii\sigma\frac{e^{(\lambda+\nu)/2}}{2}c_s^{-2}\left[(n-n_\pm)\tilde
A^{(2)l\pm1\,m}(r)+(n-n_\pm)(n-n_\pm-1)B^{(2)l\pm1\,m}(r)
\right]\nn\\
&&+\ii\sigma\frac{e^{(3\lambda-\nu)/2}}{2}\left[(n-n_\pm)\tilde
A^{(0)l\pm1\,m}(r)+(n-n_\pm)(n-n_\pm-1)B^{(0)l\pm1\,m}(r)
\right]\label{D2lm}
\end{eqnarray}
where $f,g,\xi^{(J)},\tilde\alpha^{(J)},\tilde\beta^{(J)},\eta^{(J)},
\zeta^{(J)},C^{(I)},\tilde A^{(I)},B^{(I)}$ are quantities which
depend on the perturbations $H_2^{lm}$, $K^{lm}$, etc., and which are
given in Appendix B of K1. 

At the surface of the star we compute $H_0^{lm}$ and the other
perturbations in terms of $H_2^{lm}$, $K^{lm}$, $Z_{RW}^{lm}$. We
impose the vanishing of the Lagrangian pressure perturbation (see
\cite{KojimaPTP}). This reduces the number of freely assigned
constants from three (times $L-|m|+1$), which correspond to the three
independent solutions (\ref{condcenter1})-(\ref{condcenter3}), to two
(times $L-|m|+1$), i.e. $N=2$ as discussed in Section
\ref{standingrot}.

Finally,  the equations in vacuum  are, as in K1,
\begin{eqnarray}
L_{ext}^{(J)lm}[H_0,\tilde H_1,K]
&=&\frac{\omega}{\sigma}\left(m\hat E^{(J)lm}[H_0,\tilde H_1,K]
+\frac{Q_{l-1\,m}\hat D^{(J)l-1\,m}[Z_{RW}]}{n-n_-}+
\frac{Q_{l+1\,m}\hat D^{(J)l+1\,m}[Z_{RW}]}{n-n_+}
\right)\label{outrot}\\
L^{lm}[Z_{RW}]&=&\frac{\omega}{\sigma}\left(m\hat N^{lm}[Z_{RW}]
-\frac{Q_{l-1\,m}\hat F^{l-1\,m}[H_0,K]}{n-n_-}-
\frac{Q_{l+1\,m}\hat F^{l+1\,m}[H_0,K]}{n-n_+}\right)\label{RWoutrot}
\end{eqnarray}
($J=1,\dots,4$) where $L_{ext}^{(J)lm}[H_0,\tilde H_1,K]$,
$L^{lm}[Z_{RW}]$ are the operators defined in
(\ref{in1norot})-(\ref{algebraicnorot}), (\ref{RWinnorot}), and the
expressions of $\hat E^{(J)lm}[H_0,K]$, $\hat D^{(J)lm}[Z_{RW}]$,
$\hat N^{(J)lm}[Z_{RW}]$, $\hat F^{(J)lm}[H_0,K]$ are given in K1.

When $r\gg R$, the background spacetime is with good approximation
spherically symmetric, because the terms due to rotation decrease
faster than the ``Schwarzschild-like'' components (see for instance
\cite{MTW}, Chap. 19). Therefore, spacetime perturbations satisfy the
Zerilli and the Regge-Wheeler equations.

In this limit, equation (\ref{RWoutrot}) becomes the Regge-Wheeler
equation for the function $Z_{RW}^{lm}$, whereas the Zerilli function
$Z_{Zer}^{lm}$ is related to the solution of equation (\ref{outrot})
by (\ref{defzer}).  At radial infinity, the amplitude of the
stationary wave
$\left(A_{Zer\,in}^{lm}(\sigma),A_{RW\,in}^{lm}(\sigma)\right)$ can be
computed in terms of $Z_{Zer}^{lm}$ and $Z_{RW}^{lm}$.  We can then
apply the stationary wave approach described in Section
\ref{standingwave} and in Appendix \ref{RWspher2d}.

Equations (\ref{inrotnew}), (\ref{outrot}), (\ref{RWoutrot}) can be
integrated using the spectral decomposition in Chebyshev's polynomials
as explained in Section \ref{spectral}.  There is a main difference
with respect to the example described in Appendix \ref{RWspher2d},
which refers to the axial equation for a non rotating star.  While the
matrix (\ref{matrixL}) is block-diagonal -- each block corresponding
to a value of $l$ -- the matrix representing equations
(\ref{inrotnew}), (\ref{outrot}), (\ref{RWoutrot}) presents, in
addition to the block-diagonal terms, components of order
$O(\epsilon)$, which couple $l\leftrightarrow l\pm1$.

All equations in this paper have been checked using Maple, and we have
made several cross checks in order to be sure that the Fortran
implementation of the long expressions (\ref{E1lm})-(\ref{D2lm}) are
correct.
%%%%%%%%%%%%%%%%%%%%%%%%%%%%%%%%%%%%%%%%%%%%%%%%%%%%%%%%%%%%%%%%%%%%%%%%%%%%

%%%%%%%%%%%%%%%%%%%%%%%%%%%%%%%%%%%%%%%%%%%%%%%%%%%%%%%%%%%%%%%%%%%%%%%%%%%%
\end{document}